\PassOptionsToPackage{unicode}{hyperref}
\PassOptionsToPackage{hyphens}{url}
\PassOptionsToPackage{dvipsnames,svgnames,x11names}{xcolor}
\documentclass[
  12pt]{article}

\usepackage{amsmath,amssymb}
\usepackage{iftex}
\ifPDFTeX
  \usepackage[T1]{fontenc}
  \usepackage[utf8]{inputenc}
  \usepackage{textcomp} 
\else 
  \usepackage{unicode-math}
  \defaultfontfeatures{Scale=MatchLowercase}
  \defaultfontfeatures[\rmfamily]{Ligatures=TeX,Scale=1}
\fi
\usepackage{lmodern} 
\ifPDFTeX\else  
\fi
\IfFileExists{upquote.sty}{\usepackage{upquote}}{}
\IfFileExists{microtype.sty}{
  \usepackage[]{microtype}
  \UseMicrotypeSet[protrusion]{basicmath} 
}{}
\makeatletter
\@ifundefined{KOMAClassName}{
  \IfFileExists{parskip.sty}{%
    \usepackage{parskip}
  }{
    \setlength{\parindent}{0pt}
    \setlength{\parskip}{6pt plus 2pt minus 1pt}}
}{
  \KOMAoptions{parskip=half}}
\makeatother
\usepackage{xcolor}
\usepackage[normalem]{ulem}
\makeatletter
\ifx\paragraph\undefined\else
  \let\oldparagraph\paragraph
  \renewcommand{\paragraph}{
    \@ifstar
      \xxxParagraphStar
      \xxxParagraphNoStar
  }
  \newcommand{\xxxParagraphStar}[1]{\oldparagraph*{#1}\mbox{}}
  \newcommand{\xxxParagraphNoStar}[1]{\oldparagraph{#1}\mbox{}}
\fi
\ifx\subparagraph\undefined\else
  \let\oldsubparagraph\subparagraph
  \renewcommand{\subparagraph}{
    \@ifstar
      \xxxSubParagraphStar
      \xxxSubParagraphNoStar
  }
  \newcommand{\xxxSubParagraphStar}[1]{\oldsubparagraph*{#1}\mbox{}}
  \newcommand{\xxxSubParagraphNoStar}[1]{\oldsubparagraph{#1}\mbox{}}
\fi
\makeatother

\usepackage{longtable,booktabs,array}
\usepackage{calc} 
\usepackage{etoolbox}
\makeatletter
\patchcmd\longtable{\par}{\if@noskipsec\mbox{}\fi\par}{}{}
\makeatother
\IfFileExists{footnotehyper.sty}{\usepackage{footnotehyper}}{\usepackage{footnote}}
\makesavenoteenv{longtable}
\usepackage{graphicx}
\makeatletter
\def\maxwidth{\ifdim\Gin@nat@width>\linewidth\linewidth\else\Gin@nat@width\fi}
\def\maxheight{\ifdim\Gin@nat@height>\textheight\textheight\else\Gin@nat@height\fi}
\makeatother
\setkeys{Gin}{width=\maxwidth,height=\maxheight,keepaspectratio}
\makeatletter
\def\fps@figure{htbp}
\makeatother

\makeatletter
\@ifpackageloaded{caption}{}{\usepackage{caption}}
\AtBeginDocument{%
\ifdefined\contentsname
  \renewcommand*\contentsname{Table of contents}
\else
  \newcommand\contentsname{Table of contents}
\fi
\ifdefined\listfigurename
  \renewcommand*\listfigurename{List of Figures}
\else
  \newcommand\listfigurename{List of Figures}
\fi
\ifdefined\listtablename
  \renewcommand*\listtablename{List of Tables}
\else
  \newcommand\listtablename{List of Tables}
\fi
\ifdefined\figurename
  \renewcommand*\figurename{Figure}
\else
  \newcommand\figurename{Figure}
\fi
\ifdefined\tablename
  \renewcommand*\tablename{Table}
\else
  \newcommand\tablename{Table}
\fi
}
\@ifpackageloaded{float}{}{\usepackage{float}}
\floatstyle{ruled}
\@ifundefined{c@chapter}{\newfloat{codelisting}{h}{lop}}{\newfloat{codelisting}{h}{lop}[chapter]}
\floatname{codelisting}{Listing}

\makeatother
\makeatletter
\@ifpackageloaded{caption}{}{\usepackage{caption}}
\@ifpackageloaded{subcaption}{}{\usepackage{subcaption}}
\makeatother

\ifLuaTeX
  \usepackage{selnolig}  
\fi
\usepackage[]{natbib}
\usepackage{bookmark}

\IfFileExists{xurl.sty}{\usepackage{xurl}}{} 
\hypersetup{
  pdftitle={Title},
  pdfauthor={Author 1; Author 2},
  pdfkeywords={3 to 6 keywords, that do not appear in the title},
  colorlinks=true,
  linkcolor={blue},
  filecolor={Maroon},
  citecolor={Blue},
  urlcolor={Blue},
  pdfcreator={LaTeX via pandoc}}

\def \P {\mathbb{P}}

\def \bE {\mathbb{E}}

\usepackage{arydshln}

\usepackage{amsmath}
\usepackage{amsthm}

\usepackage{hyperref}

\usepackage{listings}
\usepackage{xcolor}
\definecolor{codegreen}{rgb}{0,0.6,0}
\definecolor{codegray}{rgb}{0.5,0.5,0.5}
\definecolor{codepurple}{rgb}{0.58,0,0.82}
\definecolor{backcolour}{rgb}{0.95,0.95,0.92}

\lstdefinestyle{mystyle}{
    backgroundcolor=\color{backcolour},   
    commentstyle=\color{codegreen},
    numberstyle=\tiny\color{codegray},
    stringstyle=\color{codepurple},
    basicstyle=\ttfamily\footnotesize,
    breakatwhitespace=false,         
    breaklines=true,                 
    captionpos=b,                    
    keepspaces=true,                 
    numbers=left,                    
    numbersep=5pt,                  
    showspaces=false,                
    showstringspaces=false,
    showtabs=false,                  
    tabsize=2
}

\usepackage{tcolorbox}

\newcommand{\anon}{1}

\begin{document}

\def\spacingset#1{\renewcommand{\baselinestretch}%
{#1}\small\normalsize} \spacingset{1}


\if1\anon
{
  \title{\bf \Large Adaptive Influence-Based Borrowing Framework for Improving Treatment Effect Estimation in RCTs Using External Controls} 
  \author{Peng Wu,  Jile Chaoge \\
  School of Mathematics and Statistics, \\
  Beijing Technology and Business University\\
  \\
    Shu Yang\thanks{Corresponding Author. \\
    }\hspace{.2cm} \\
    Department of Statistics, North Carolina State University}
  \date{}
  \maketitle
} \fi

\if0\anon
{
  \title{\bf \Large } 
  \bigskip
  \bigskip
  \bigskip
  \begin{center}
    {\LARGE\bf Adaptive Influence-Based Borrowing Framework for Improving Treatment Effect Estimation in RCTs Using External Controls}
\end{center}
  \medskip
} \fi

\begin{abstract}
Randomized controlled trials (RCTs) often suffer from limited sample sizes due to high
costs and lengthy recruitment periods, compromising precision in treatment effect estimation. External real-world control data offer a valuable opportunity for augmentation, but naïve integration may introduce bias without careful compatibility assessment. This paper presents a practical tutorial on the adaptive influence-based borrowing framework~\citep{Yang-etal2026}, which addresses this challenge through a principled, individual-level borrowing strategy. The core intuition is straightforward: rather than indiscriminately pooling all external controls (ECs), the framework first asks how much each external patient would perturb the outcome model fitted using RCT controls. External patients whose inclusion barely changes this model are deemed comparable and prioritized for borrowing, whereas those who substantially shift it are flagged as potentially incompatible. This individual-level compatibility metric, based on the influence score, is then used to construct a sequence of nested candidate subsets of ECs, from which the optimal subset is selected by minimizing the mean squared error of the treatment effect estimator, balancing the competing risks of bias from over-borrowing and imprecision from under-borrowing. When systematic differences between ECs and RCT controls are substantial, an optional outcome calibration step can align the two groups before influence-based selection proceeds. We provide a clear, step-by-step workflow  with emphasis on methodological intuition, practical considerations, and visualization, thereby offering a principled, transparent, and practical method for leveraging ECs when RCTs alone are underpowered. Implementation is supported by an accompanying \texttt{R} package {\bf InfluenceBorrowing}.  
\end{abstract}

\noindent%
{\it Keywords:} Adaptive Borrowing, Causal Inference, Data Integration, Efficiency Improvement. 
\vfill

\newpage
\spacingset{1.8} 

\section{Introduction}\label{sec-intro} 
RCTs are the gold standard for evaluating treatment efficacy because randomization eliminates confounding and supports internally valid causal inference~\citep{Imbens-Rubin2015, Hernan-Robins2020}. However, conducting adequately powered RCTs is often logistically demanding and costly. In rare diseases and areas of high unmet need, the eligible patient pool may be too small to support a conventionally powered trial, and randomly assigning patients to placebo or inferior controls raises serious ethical concerns that can deter enrollment. These practical and ethical constraints routinely result in small sample sizes that, while internally valid, yield imprecise treatment effect estimates with wide confidence intervals and limited statistical power~\citep{Gao2024Adaptive}. At the same time, the rapid growth of real-world data, including disease registries, electronic health records, and historical clinical trials, creates new opportunities to augment RCTs~\citep{Qiu-etal2015, Colnet-etal2024,  Wu-etal-2025-Compare}. The U.S. Food and Drug Administration (FDA) has issued guidance encouraging the rigorous use of real-world evidence to support regulatory decision-making~\citep{FDA2021, FDA2023}. Hybrid control arm designs, which combine the current RCT with external controls (ECs), are increasingly used in oncology and rare disease development programs. These borrowing approaches aim to increase effective sample size, shorten development timelines, and improve treatment effect estimation precision~\citep{Viele2014,Schmidli2014, Laan-etal2025}.

Despite their promise, ECs are rarely a perfect substitute for RCT controls, even when sponsors invest considerable effort in identifying fit-for-purpose external data sources. Careful curation of EC datasets can mitigate many obvious sources of incompatibility, such as mismatched eligibility criteria or poorly overlapping covariate distributions, yet hidden biases often persist. Unmeasured prognostic factors, temporal shifts in standard of care, site-level variability, and differences in outcome measurement practices can all introduce systematic discrepancies that observed covariates alone cannot fully capture or correct. Most existing borrowing strategies rely, explicitly or implicitly, on the exchangeability assumption: conditional on observed covariates, the distribution of potential outcomes under control is identical between RCT controls and ECs ~\citep{dahabreh2019generalizing, Dahabreh-etal2020}. In practice, this assumption is difficult to verify and frequently violated, even after rigorous data curation, and under such violations, naïve or full borrowing can introduce substantial bias and lead to misleading conclusions~\citep{Gao-etalSurvival2025, Zhu-etal-ICML25, Yang2025Influence}. 

Typically, ECs have substantially larger sample sizes than RCTs and tend to exhibit greater individual heterogeneity~\citep{li2023improving, FDA2023}. As a result, some external patients may be comparable to RCT controls, whereas others may not. This heterogeneity creates a fundamental tension between over-borrowing, which incorporates non-comparable ECs and introduces bias, and under-borrowing, which excludes useful ECs and limits efficiency gains. The central methodological questions can therefore be summarized as follows:

\begin{tcolorbox}

\textbf{Q1}: How to quantify the comparability of each EC for treatment effect estimation in RCTs, thereby identifying the potentially useful samples from the full set of ECs?

\textbf{Q2}:  How can we determine the optimal subset of comparable samples in ECs that achieves a proper balance between over-borrowing and under-borrowing?  

\textbf{Q3}: When only a small fraction of ECs are deemed comparable, can we expand the usable pool through outcome calibration?  
\end{tcolorbox}

\begin{figure}[h]
	  \vspace{8pt}
    \centering
    \includegraphics[width=2\textwidth, height=0.5\textheight, keepaspectratio]{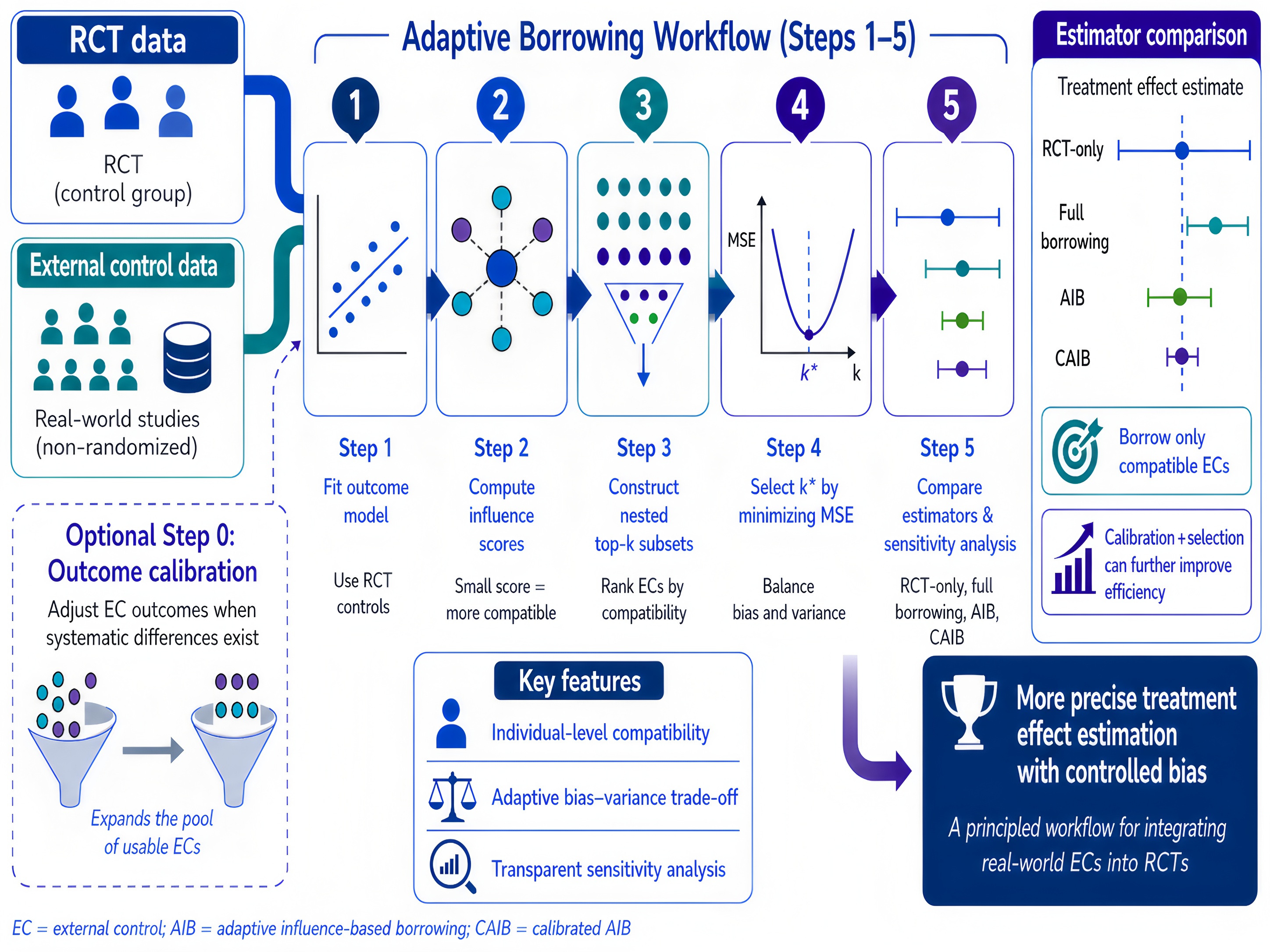}
    \caption{Workflow of adaptive influence-based borrowing framework.} 
    \label{fig-workflow}
    \vspace{-18pt}
\end{figure}

Several methods determine the degree of borrowing or adjust EC outcomes based on observed discrepancies from RCT controls~\citep{stuart2008matching, neuenschwander2009note, hobbs2011hierarchical, schoenfeld2019design}. For comprehensive discussions, see \citet{shan2022simulation} and \citet{Gao2024Adaptive}. 
\citet{Yang-etal2026} recently proposed the adaptive influence-based borrowing (AIB) framework to address this challenge. Rather than applying a single global borrowing weight or threshold, the AIB framework evaluates the compatibility of each EC individually, using influence scores to measure how much each EC perturbs the outcome model estimated from RCT controls. ECs with small influence scores, those whose inclusion barely changes the model, are identified as comparable and prioritized for borrowing. ECs with large influence scores signal potential incompatibility and are set aside (addressing Q1). An optimal subset of ECs is then selected by minimizing the mean squared error (MSE) of the combined treatment effect estimator, directly balancing bias and variance (addressing Q2). When most ECs are systematically different from RCT controls, e.g., due to a shift in standard of care across time periods, an optional outcome calibration step can first correct these systematic discrepancies before influence-based selection proceeds, expanding the pool of usable ECs (addressing Q3). 
Figure \ref{fig-workflow} summarizes the workflow of the AIB framework. 

The purpose of this paper is to serve as a practical tutorial for the AIB framework. While \citet{Yang-etal2026} establishes the theoretical foundations, the goal here is different: we aim to make the methodology accessible and immediately actionable for applied statisticians and clinical researchers. To this end, we provide a step-by-step workflow that walks through each stage of the analysis---from data preparation and outcome model fitting, through influence score computation and subset construction, to optimal selection, treatment effect estimation, and sensitivity analysis. Throughout, we emphasize methodological intuition over technical derivation, accompanying each step with reproducible R code via the companion package {\bf InfluenceBorrowing}, available on the Comprehensive R Archive Network (CRAN) at  \url{https://CRAN.R-project.org/package=InfluenceBorrowing}. Synthesized and real-world examples  illustrate the complete pipeline from raw data to a regulatory-ready result. 

The tutorial is organized as follows. Section \ref{sec2} describes the problem setting and introduces notation. Section \ref{sec3} discusses the exchangeability assumption and the consequences of its violation, providing the conceptual motivation for adaptive borrowing. Section \ref{sec4} presents the step-by-step implementation guide for the core AIB workflow, including comparisons with benchmark estimators and sensitivity analyses. Section \ref{sec5} describes the optional outcome calibration procedure for settings where systematic differences between ECs and RCT controls are widespread. Section \ref{sec6} applies the AIB framework to real-world data. Section \ref{sec7} concludes with practical guidance and a discussion of limitations.


\section{Problem Description}  \label{sec2}

In practice, RCTs may be limited in two distinct ways: the overall sample size may be small due to logistical, financial, and ethical constraints, or the control arm may be deliberately undersized to minimize the number of patients receiving no active treatment, a common design choice in rare diseases and high unmet need settings. In either case, the result is an imprecise estimation of the treatment effect, with wide confidence intervals that are uninformative for clinical or regulatory decision-making. EC data can play a valuable role in both scenarios: by supplementing the RCT control arm with comparable external patients, we can improve estimation efficiency without requiring additional randomized enrollment.

 The primary objective of the AIB framework is to enhance estimation efficiency in the RCT by leveraging ECs in a principled and data-adaptive manner.   
Formally,  let $X$ denote the observed pre-treatment covariates, and $A \in \{0, 1\}$ denote the binary treatment assignment, where $A=1$ indicates treatment and $A=0$ indicates control. Let $Y$  denote the outcome of interest. Under the potential outcomes framework~\citep{rubin1974estimating,splawa1990application} in causal inference, let $Y(1)$ and $Y(0)$ denote the potential outcomes that would be observed if the individual were assigned the treatment  ($A=1$) and control ($A=0$), respectively.   The observed outcome equals the potential outcome under the treatment actually assigned, i.e., $Y = AY(1) + (1 - A)Y(0)$.  
Suppose we have access to RCT data and EC data, denoted by 
 \begin{equation*}
      \mathcal{D}_{\text{RCT}} = \{(X_i, A_i, Y_i),  i = 1, ..., n_1 \}, \qquad   \mathcal{D}_{\text{EC}} = \{(X_i, A_i = 0, Y_i),  i = n_1 +1, ..., n_1 + n_0\}. 
\end{equation*} 
\begin{figure}[h]
\vspace{-16pt}
    \centering
    \includegraphics[width=0.85\textwidth]{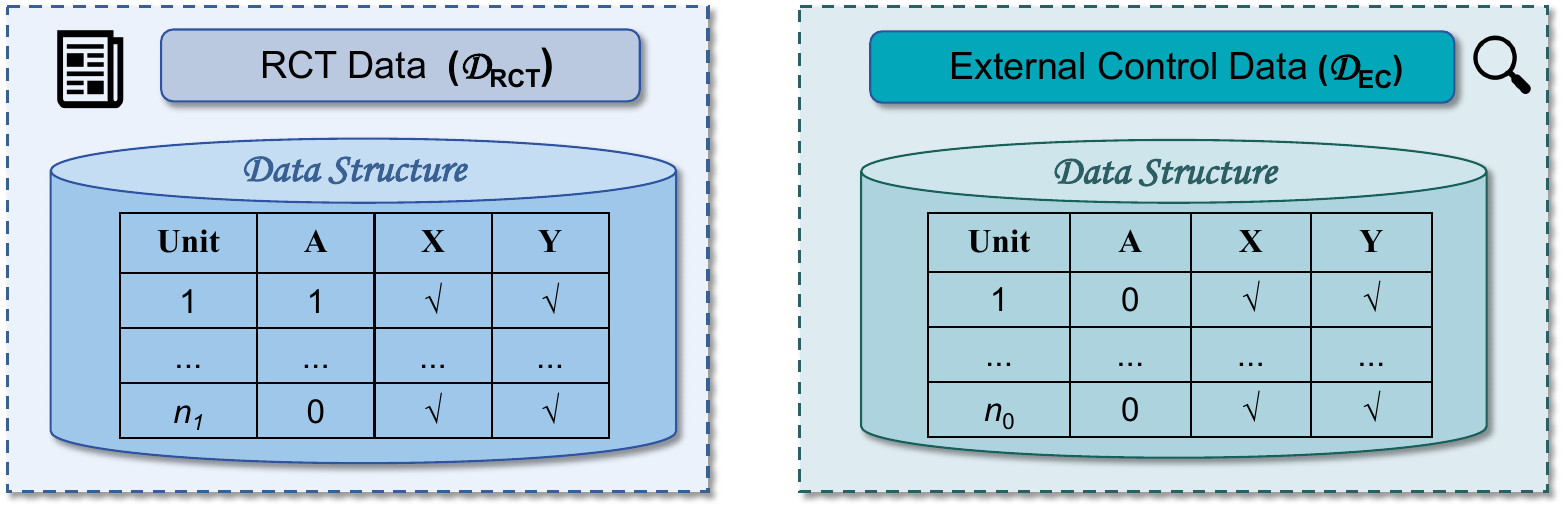}
    \caption{Observed data} 
    \label{fig1} 
    \vspace{-18pt}
\end{figure}

The RCT data contains $n_1$ units, the EC data contains $n_0$ units with only controls ($A_i$ = 0 for all ECs). The observed data structures are shown in Figure \ref{fig1}. Suppose that $\mathcal{D}_{\text{RCT}}$ and $\mathcal{D}_{\text{EC}}$ consist of independent and identically distributed samples drawn from the probability distributions  $\P$ and $\mathbb{Q}$, respectively. 
The target causal estimand is the average treatment effect  in the RCT population, defined as $$\tau=\mathbb{E}_{\P}[Y(1)-Y(0)],$$
  where $\mathbb{E}_{\P}$ denotes taking expectation with respect to $\P$.   
This estimand represents the average causal effect of treatment versus control for the population that satisfies the RCT eligibility criteria.    
Because treatment assignment is randomized in the RCT, $A$ is independent of potential outcomes $(Y(1), Y(0))$ given covariates $X$, ensuring internal validity of causal comparisons within the trial. 
  Thus, we could  obtain consistent estimators of $\tau$ using the RCT data alone, see Section \ref{sec4-6-1} for more details. 
In our setting, the target population remains that of the RCT population. The role of the EC data is not to redefine the estimand, but rather to improve the estimation efficiency of $\tau$.

\section{Why Na\"{i}ve Borrowing Can Go Wrong: Exchangeability and Its Violations}  \label{sec3}

Borrowing ECs can substantially improve precision when appropriately implemented \citep{Viele2014,Schmidli2014}. However, simply pooling ECs with RCT controls without
careful assessment of comparability can introduce bias and compromise the validity of treatment effect estimation \citep{Gao2024Adaptive}. 
In this section, we explain the key exchangeability assumption commonly used in external borrowing methods, discuss how it may fail in practice, and  the consequences of its violation. 

{\bf Exchangeability Assumption}.  A central assumption for valid EC augmentation is exchangeability between RCT controls and ECs~\citep{Yang2025Influence}: 
\begin{equation}
          \P( Y(0) \mid X)  =   \mathbb{Q}( Y(0) \mid X  ). 
\end{equation}  
It states that, conditional on baseline covariates, control outcomes have the same distribution in RCT controls and ECs. In other words, after covariate adjustment, EC units are comparable to RCT controls, with no residual systematic differences in prognosis. When this assumption holds, ECs can validly contribute information about the control outcome distribution in the RCT population.


For clinicians, exchangeability requires that external patients reflect the same underlying disease process and standard-of-care experience as RCT controls. This involves ensuring similar inclusion and exclusion criteria, comparable background therapies, consistent outcome definitions and follow-up schedules, and treatment within the same clinical era. These considerations are emphasized in regulatory guidance on the use of real-world evidence and ECs~\citep{FDA2023,EMA2023}. When these aspects are broadly aligned, the exchgeability assumption may be reasonable. 

{\bf Common Sources of Exchangeability Violations}.  
In practice, exchangeability is often hard to satisfy, and several common factors can lead to its violation:  
\begin{itemize}
\item Differences in standard-of-care. Clinical practice evolves over time. Improvements in supportive care, diagnostic tools, or background therapies change patient outcomes. 
If EC participants received care under a different standard than RCT controls, their outcomes may differ systematically, even after covariate adjustment~\citep{Viele2014}.

\item Calendar time effects. ECs may come from earlier or later periods than the RCT. Secular trends in disease management, healthcare access, or patient demographics can introduce systematic differences unrelated to treatment assignment~\citep{FDA2023}.  

\item Measurement heterogeneity. Differences in outcome definitions, assessment frequency, imaging technology, or adjudication procedures can affect observed outcomes. Outcomes measured in routine practice may not be directly comparable to those assessed under strict trial protocols~\citep{Sherman2016}.  


\item Unmeasured prognostic factors. Even when measured covariates appear similar, unobserved factors—such as disease severity markers, frailty, socioeconomic status, or adherence patterns—may differ between data sources. These residual differences cannot be removed through statistical adjustment alone~\citep{Stuart2010}.  
\end{itemize}

A critical point for practitioners is that these sources of violation are often hidden: they may not be apparent from covariate distributions alone, and standard balance diagnostics may fail to detect them. This is precisely what makes na\"{i}ve borrowing risky even after careful data curation.

{\bf Consequences of Violation}.  
Failure of exchangeability has important statistical and clinical consequences: (a) Biased treatment effect estimates.   
    If ECs systematically differ from RCT controls after adjusting for covariates, including  them may distort the estimation of treatment effects. 
     (b) Misleading clinical conclusions.   
    Biased treatment effect estimates may lead to incorrect inferences regarding efficacy. Overestimation of benefit may expose patients to ineffective or harmful treatments, whereas underestimation may delay access to beneficial therapies; (c) Reduced regulatory credibility.   
    Regulatory agencies carefully scrutinize externally augmented analyses. If comparability between data sources is not convincingly demonstrated, augmented analyses may be discounted or rejected~\citep{FDA2023,EMA2023}.  
%
%
For these reasons, EC augmentation should not be treated as a simple data pooling exercise. Instead, it requires careful assessment of compatibility between RCT controls and ECs, along with methods that adaptively limit borrowing when exchangeability is questionable \citep{Schmidli2014}. 

\section{Step-by-Step Implementation Guide}  \label{sec4}

In this section, we present a step-by-step roadmap for implementing the AIB framework, with detailed explanations of its key ideas. The workflow proceeds from outcome calibration to optimal subset selection and final treatment effect estimation.


\subsection{Overall Workflow} \label{sec4-1}
The workflow for the AIB framework consists of several steps:

\quad
 {\bf Step 0 (Optional --- Outcome calibration)}: 
 When systematic differences between ECs and RCT controls are pervasive, adjust EC outcomes before proceeding. This step is described in detail in Section \ref{sec5} and is optional if the two groups appear broadly comparable.

 \quad
 {\bf Step 1 (Outcome model fitting on RCT controls)}:  
Fit an outcome regression model using RCT controls only, capturing the relationship between baseline covariates and the control outcome within the RCT population.
 
  \quad
 {\bf Step 2 (Influence score calculation)}: For each EC unit, compute an influence score quantifying how much its inclusion would perturb the outcome model fitted in Step 1. This provides an individual-level measure of compatibility with the RCT control arm.

   \quad
 {\bf Step 3 (Candidate borrowing subsets construction)}:  
Based on the influence scores, EC units are ranked, and top-$k$ nested subsets are constructed by sequentially adding those with the smallest scores, yielding a sequence of candidate borrowing subsets.

    \quad
 {\bf Step 4 (Optimal subset selection and treatment effect estimation)}: For each candidate subset, incorporate it into the estimation of $\tau$ and estimate the associated bias and variance. Then compute the MSE and select the subset that minimizes it. Finally, return the estimate of $\tau$ corresponding to the selected subset of ECs.

    \quad
 {\bf Step 5 (Comparison of Estimators and Sensitivity Analysis)}: We compare the proposed estimator from Step 4 with two RCT-only estimators (see Section \ref{sec4-6}), as well as with the full-borrowing estimator that leverages RCT data and all ECs. Additionally, we conduct sensitivity analyses to assess the impact of varying the    top-$k$ ECs and alternative specifications of the nuisance models.


The \texttt{R} package {\bf InfluenceBorrowing} facilitates implementing the AIB  framework. We can install this package in CRAN by using the following standard code. 
\begin{lstlisting}[language=R]
install.packages("InfluenceBorrowing") 
\end{lstlisting}

Before detailing each step, we first describe the data preparation process. Suppose we have access to both RCT and EC data. We use a simulated example to illustrate the procedure.

\begin{lstlisting}[language=R]
library(InfluenceBorrowing)
n_rct <- 100  # sample size of RCT data
n_ec <- 200   # sample size of EC data
Dat <- gen_demo_data(n_rct = 100, n_ec = 200)     
data_rct <- Dat$data_rct                 # the simulated RCT data 
data_ec  <- Dat$data_ec                  # the simulated EC data 
ATE_true <- Dat$ATE_true                 # the true value of ATE in RCT 
data_rct_control <- data_rct[data_rct$A == 0, ]  # RCT controls
data_rct_control$A <- NULL          # remove the treatment 
\end{lstlisting}

\begin{figure}[t]
\centering
\vspace{2pt} 
\subfloat[]{
\begin{minipage}[t]{0.5\linewidth}
\centering
\includegraphics[width=1\textwidth]{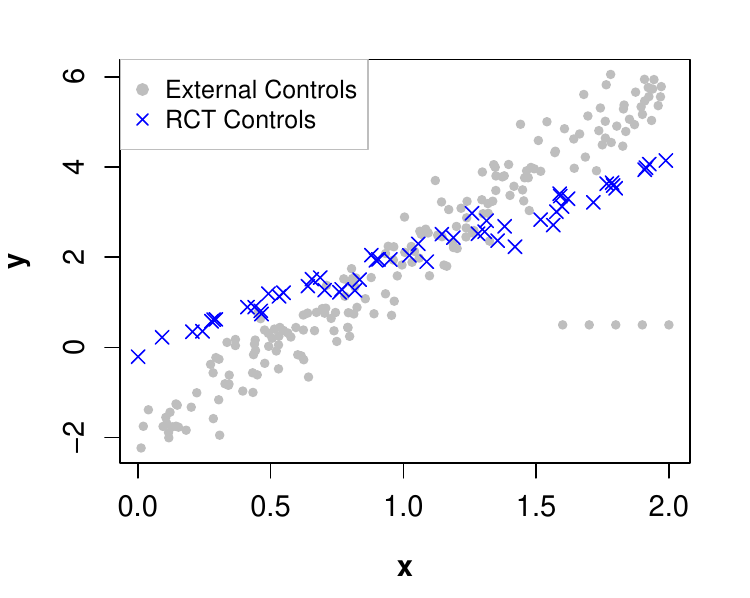}
\end{minipage}%
} 
\subfloat[]{
\begin{minipage}[t]{0.5\linewidth}
\centering
\includegraphics[width=1\textwidth]{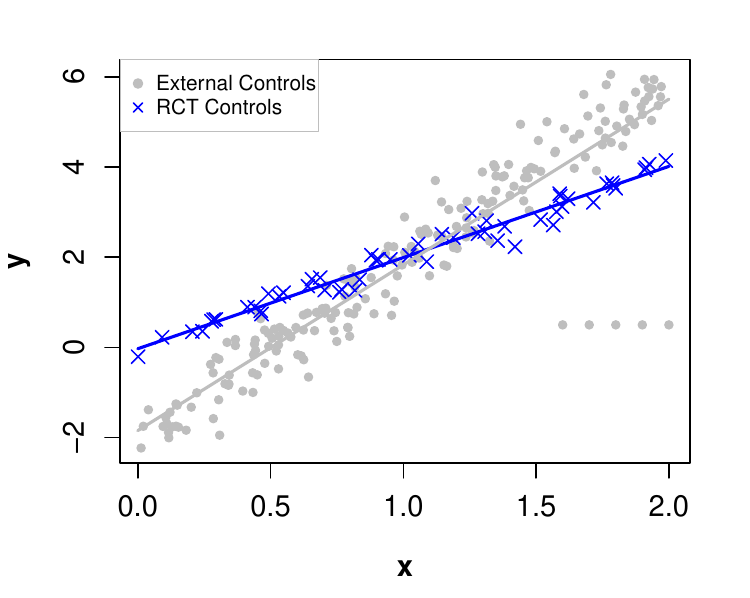}
\end{minipage}%
}%
\vspace{-10pt}
\caption{(a) Scatter plot for ECs and RCT controls; (b) Fitted linear model on RCT controls (blue line) and on ECs (gray line).}
\label{fig3}
\end{figure}

We present a scatter plot of ECs and RCT controls in Figure \ref{fig3}(a). The ECs appear markedly different from the RCT controls, with only limited overlap between them. 

\subsection{Step 1: Fit Outcome Model on RCT Controls}

The first step is to model the relationship between covariates \(X\) and the outcome \(Y\) using only RCT controls. That is, we aim to estimate 
		\[       \mu_0(x) = \bE_{\P}(Y\mid X=x, A = 0).  \]
Suppose $\mu_0(x)$ is modeled as $\mu_0(x; \theta)$ with parameter $\theta$. For continuous outcomes, we may use a linear model, $\mu_0(x; \theta) = x^\intercal \theta$; For binary outcomes, we may use a logistic regression model,  $\mu_0(x; \theta) = \exp(x^\intercal \theta)/\{1 + \exp(x^\intercal \theta)\}$. 
Currently, {\bf InfluenceBorrowing} supports generalized linear models and implements them via the \texttt{glm()} function.

\begin{lstlisting}[language=R]
# fit linear model on RCT controls     
model_mu0 <- glm(Y ~ X, data = data_rct_control,    
             family = gaussian(link = "identity"))  
\end{lstlisting} 
 Let $\hat{\theta}$ denote the estimated parameter, and define 
 $$\hat \mu_0(x) = \mu_0(x; \hat \theta)$$   
as the fitted value of $\mu_0(x)$. The fitted model is shown in Figure \ref{fig3}(b) (blue line), along with the model fitted on ECs (gray line) for comparison.

\subsection{Step 2: Calculate  Influence Score for Each External Control}

With the outcome model in hand, we now address the first key methodological question: 
\begin{tcolorbox}
\textbf{Q1}: How to quantify the comparability of each EC for treatment effect estimation in RCTs, thereby identifying the potentially useful samples from the full set of ECs?
\end{tcolorbox}
The AIB framework answers this using influence scores, a classical tool from robust statistics  \citep{cook1980characterizations, koh2017understanding}  that measures how much a single observation perturbs a fitted model. The intuition is direct: an EC unit whose covariate-outcome pattern is consistent with the RCT control arm will have little effect on the fitted model when added to RCT controls; one that is systematically different will shift the model noticeably. 

{\bf Influence Score.} The AIB framework adopts influence scores to quantify the comparability of each EC. Specifically, the estimator of  $\theta$ is defined as: 
\begin{equation}\label{erm} 
    \hat{\theta}\overset{\text{def}}{=} \arg\min_{\theta \in \Theta} \sum_{Z_i \in \mathcal{C}} L(Z_i;\theta),
\end{equation}
where $Z_i \overset{\text{def}}{=} (X_i, Y_i)$, $L(Z_i; \theta)$ is a loss function that is twice-differentiable and convex in $\theta$, and $C \overset{\text{def}}{=} \{(X_i, A_i = 0, Y_i) \in \mathcal{D}_{\text{RCT}}\}$ is RCT controls.  
For any given EC unit $z=(x,y)$, let $\hat{\theta}_{+z}$ denote the modified parameter  by refitting the model after adding $z$ to RCT controls: 
$$
    \hat{\theta}_{+z}\overset{\textup{def}}{=}\arg\min_{\theta\in\Theta} \sum_{(X_i, Y_i) \in\mathcal{C}\cup z} L(Z; \theta).
$$ 
The influence of the EC unit $z$ on the loss over RCT controls can then be measured as
\[ 
  \sum_{Z_i\in\mathcal{C}}|L(Z_i,\hat{\theta}_{+z})-L(Z_i,\hat{\theta})|,
  \]
where a larger value indicates that $z$ has a greater impact on $\hat{\mu}_0(x) = \mu_0(x; \hat{\theta})$.   
However, refitting the model for each added unit $z$ is computationally expensive. 
Fortunately,  following \citet{cook1980characterizations} and \citet{koh2017understanding}, we could approximate it using 
       \begin{equation}  \label{eq3}
        \mathcal{IF}(z)
        \overset{\text{def}}{=}\sum_{Z_i\in\mathcal{C}}\left|\nabla_\theta L(Z_i,\hat{\theta})^\top H_{\hat{\theta}}^{-1} \nabla_\theta L(z, \hat{\theta})\right|,  
    \end{equation}
   where $H_{\hat{\theta}}\overset{\mathrm{def}}{=} |\mathcal{C}|^{-1} \sum_{Z_i\in\mathcal{C}} \nabla_{\hat{\theta}}^{2} L(Z_{i}, \hat{\theta})$ is the Hessian matrix, $\nabla_\theta L(z, \hat{\theta})$ is the gradient of $L(z, \hat{\theta})$ with respect to $\theta$, evaluated at $\theta = \hat \theta$. 
From equation \eqref{eq3}, to calculate $\mathcal{IF}(z)$ for each $z$ in ECs, the terms $\nabla_\theta L(Z_i, \hat{\theta})$ and $H_{\hat{\theta}}$ are identical across different $z$ and need to be computed only once, avoiding the need to repeatedly refit models for each EC.

\textbf{Intuitive Interpretation and Strengths}. Intuitively, an EC whose covariate-outcome relationship aligns closely with that of RCT controls will exert limited influence on the estimated model and can be safely borrowed. Conversely, an EC that substantially shifts the model may indicate structural differences and an increased risk of bias, signaling potential non-comparability.  
We can interpret influence scores as individualized comparability metrics:
  \begin{itemize}
\item Small influence score $\longrightarrow$ the EC unit reflects a similar disease process, treatment experience, and outcome behavior to RCT controls, and is a strong candidate for borrowing. 
\item Large influence score $\longrightarrow$ the EC unit may differ from RCT controls in unmeasured prognostic factors, calendar effects, or standard-of-care practices. 
\end{itemize}
Notably, the influence score is robust to outliers in ECs. It is defined at the individual level, with each EC's score computed independently of all other ECs, making the approach inherently robust to outliers. This is a key strength, as ECs are typically drawn from large and heterogeneous sources~\citep{li2023improving, Colnet-etal2024}, making it inevitable that some EC units (e.g., outliers) exhibit patterns that differ substantially from those of RCT controls. 
In addition, calculating influence scores for ECs does not require modeling the outcome regression in ECs, which is another key strength.

{\bf Implementation.} We compute influence scores using the \texttt{compute\_influences()} function. 
  
\begin{figure}[h]
\vspace{-16pt}
    \centering
    \includegraphics[width=0.9\textwidth]{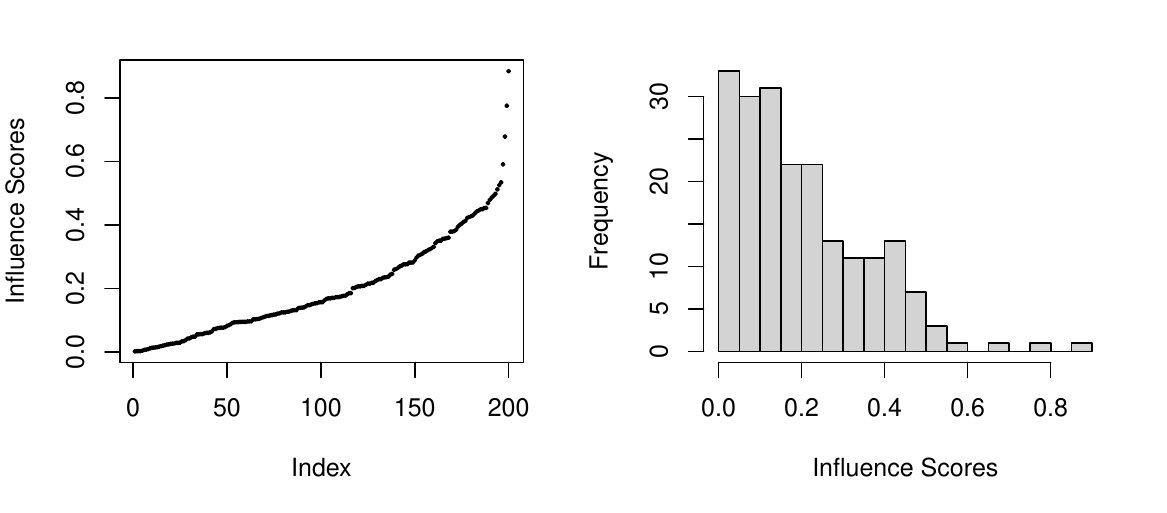}
    \vspace{-16pt}
    \caption{Sorted values and histogram of influence scores} 
    \label{fig5} 
\end{figure}

\begin{lstlisting}[language=R]
data_ec_xy <- data.frame(X = data_ec$X, Y = data_ec$Y)
influences <- compute_influences(model_mu0, testdata = data_ec_xy) 
\end{lstlisting}
We present the sorted values and histogram of the influence scores in Figure \ref{fig5}. It indicates that several ECs have small influence scores and are suitable candidates for borrowing. 

\subsection{Step 3: Construct Nested Candidate Borrowing Subsets}

Based on the influence scores, we rank all ECs in ascending order, following the principle that smaller scores indicate greater comparability. This ranking naturally defines a sequence of nested candidate subsets  \(\mathcal{S}_1 \subset \mathcal{S}_2 \subset \dots \subset \mathcal{S}_{n_0}\), where \(\mathcal{S}_k\) contains the top-\(k\) ECs with the smallest influence scores, $n_0$ is the total number of ECs. This sequence represents a continuum of borrowing, ranging from the single most comparable external control to the entire external dataset.  
For the simulated example above, we present the top 50, 75, 100, and 150 ECs, respectively, to illustrate how ECs are gradually selected. 

\begin{lstlisting}[language=R]
sorted_indices <- order(influences, decreasing = FALSE) 
top_k_list <- c(50, 75, 100, 150)
par(mfrow = c(2, 2))
for(num in top_k_list){
  selected_indices <- sorted_indices[1:num]
  data_ec_selected <- data_ec[selected_indices, ] 
  plot(data_ec$X, data_ec$Y, col = "gray", pch = 19, cex = 0.7,
       xlab = paste0('top ', num), ylab = "", font.lab = 2)
  points(data_rct_control$X, data_rct_control$Y, col = "blue", pch = 4)
  points(data_ec_selected$X, data_ec_selected$Y, col = "red", pch = 19)
  legend("topleft", legend = c("ECs", "RCT", "Selected ECs"),
         col = c("gray", "blue", "red"), pch = c(19, 4, 19), cex = 0.7)
}
\end{lstlisting}

 \begin{figure}[h]
\vspace{-16pt}
    \centering
    \includegraphics[width=0.95\textwidth]{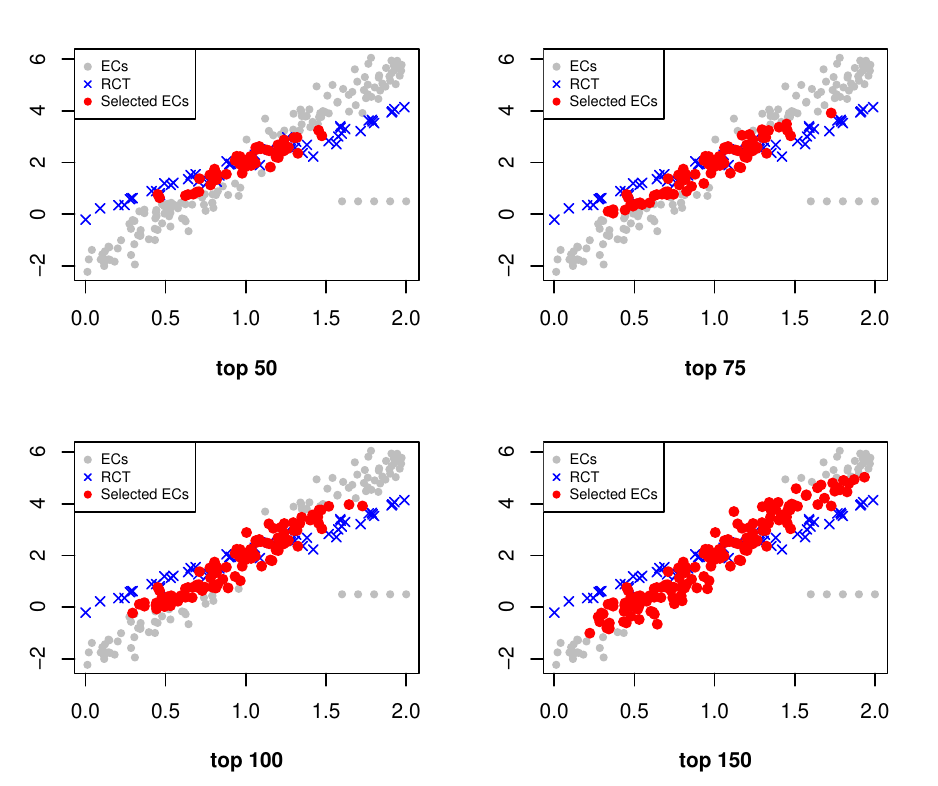}
    \vspace{-16pt}
    \caption{Top-$k$ ($k=50, 75, 100, 150$) ECs with smallest influence scores (marked in red).} 
    \label{fig6} 
\end{figure}

Figure \ref{fig6} illustrates the bias--variance trade-off as the number of selected ECs ($k$) increases: 
\begin{itemize}
    \item For small $k$ (e.g., $k \leq 50$), the selected ECs are highly comparable to RCT controls. Borrowing them primarily increases the effective sample size of controls, which helps reduce variance, with minimal concern for bias.
    
    \item As $k$ increases further, including additional ECs may introduce non-comparable samples, resulting in greater bias.
    
\end{itemize} 
Therefore, fully utilizing the ECs requires striking an appropriate balance between bias and variance. In the next subsection, we achieve this by minimizing MSE.

\subsection{Step 4: Select the Optimal Subset and Obtain Final Treatment Effect Estimation}

We now turn to address Q2. 
\begin{tcolorbox}
\textbf{Q2}:  How can we determine the optimal subset of comparable samples in ECs that achieves a proper balance between over-borrowing and under-borrowing?  
\end{tcolorbox}

The AIB framework proposes to find the candidate subset \(\mathcal{S}_k\) that minimizes the MSE of the average treatment effect estimator.  
The procedures are given below.  
\begin{itemize} 
    \item \emph{Estimate $\tau$ using  \(\mathcal{S}_k \cup \mathcal{D}_{\text{RCT}}\) (for $k =1, ..., n_0$)}. For each candidate subset $\mathcal{S}_k$, we estimate $\tau$ by combining it with the RCT data ($\mathcal{D}_{\text{RCT}}$).  For ease of presentation, we denote $\P_{\mathcal{S}_k}$ as the combined population of  RCT data and the selected top-$k$ ECs.
    The estimator of $\tau$ is given as 
    	\begin{equation}  \label{eq-AIB}       \hat{\tau}_{\mathcal{S}_k} = \frac{1}{n_0 + k} \sum_{i \in \mathcal{S}_k \cup \mathcal{D}_{\text{RCT}}}  \varphi( X_i, A_i, Y_i; \hat \eta),  \end{equation}
where 
     \begin{align*}   
     \varphi( X_i, A_i, Y_i; \hat \eta) =   \frac{ A_i(Y_i - \hat m_{\mathcal{S}_k, 1}(X_i))}{ \hat e_{\mathcal{S}_k}(X_i)  }   -   \frac{(1-A_i)(Y_i - \hat m_{\mathcal{S}_k, 0}(X_i)) }{1 - \hat e_{\mathcal{S}_k}(X_i)  }  +  \hat m_{\mathcal{S}_k, 1}(X_i) - \hat m_{\mathcal{S}_k, 0}(X_i),  
  \end{align*}
and  $\hat{\eta} =  (\hat e_{\mathcal{S}_k}(x), \hat m_{\mathcal{S}_k, 1}(x), \hat m_{\mathcal{S}_k, 0}(x))$ 
  is the estimate of nuisance parameter vector $\eta = ( e_{\mathcal{S}_k}(x), m_{\mathcal{S}_k, 1}(x), m_{\mathcal{S}_k, 0}(x))$, defined as follows: $e_{\mathcal{S}_k}(x) = \P_{\mathcal{S}_k}(A=1 \mid X=x)$ is the propensity score in the combined population, $m_{\mathcal{S}_k, a}(x) = \bE_{\P_{\mathcal{S}_k}} [Y \mid X=x, A=a]$ for $a = 0, 1$ are the outcome regression functions in the combined population. 

   \item \emph{Calculate the MSE of $\hat{\tau}_{\mathcal{S}_k}$}. The estimated bias and variance of $\hat{\tau}_{\mathcal{S}_k}$ are given by  
    \[
    \widehat{\text{bias}}(\hat{\tau}_{\mathcal{S}_k}) = \hat{\tau}_{\mathcal{S}_k} - \tau_{\text{reference}},
    \]
  and  $ \widehat{\text{var}}(\hat{\tau}_{\mathcal{S}_k})$ is the sample variance of $\{  \varphi( X_i, A_i, Y_i; \hat \eta), i \in \mathcal{S}_k \cup \mathcal{D}_{\text{RCT}}\}$.  
  The MSE is 
    \[
    \widehat{\text{MSE}}(\hat{\tau}_{\mathcal{S}_k}) = \widehat{\text{bias}}(\hat{\tau}_{\mathcal{S}_k})^2 + \widehat{\text{var}}(\hat{\tau}_{\mathcal{S}_k}).
    \]
  The $\tau_{\text{reference}}$ denotes the reference value of $\tau$. In simulations, $\tau_{\text{reference}}$ is set to the true value of $\tau$; in real-world applications, we use the augmented inverse probability weighting (AIPW) estimator based solely on the RCT data~\citep{Bang-Robins-2005}: 
  	\begin{equation}   \label{eq-AIPW}   \hat \tau_{\textup{aipw}} = \frac{1}{n_1} \sum_{i\in \mathcal{D}_{\text{RCT}} } \biggl\{\frac{ A_i(Y_i - \hat \mu_1(X_i))}{\hat e_1(X_i)} - \frac{(1-A_i)(Y_i - \hat \mu_0(X_i))}{1 - \hat e_1(X_i)}  + (\hat \mu_1(X_i) - \hat \mu_0(X_i))\biggr\},  \end{equation}
  where $e_1(x) = \P(A=1\mid X=x)$ is the propensity score in the RCT data, and $\mu_a(x) = \bE_{\P}(Y\mid A=a, X=x)$ for $a = 0, 1$ are the outcome regression functions in the RCT data.  
  When we take $\hat \tau_{\textup{aipw}} $ as the reference value, we actually treat it as the baseline estimator with zero bias and use ECs solely to reduce its variance. 
  In practice, $e_1(x)$ may be known a priori~\citep{Gao2024Adaptive, Qiu-etal2015}. In such a case, we set $\hat e_1(x) = e_1(x)$.
  

    \item \emph{Select the optimal subset}.  Choose \(k\) to minimize the estimated MSE: 
    \[
    k^* = \arg\min_{k} \widehat{\text{MSE}}(\hat{\tau}_{\mathcal{S}_k}).
    \]
The subset \(\mathcal{S}_{k^*}\) is selected as the optimal set of ECs for borrowing. The final estimate,
   \begin{equation}  \label{eq6}
    \hat \tau_{\text{aib}}  = \hat{\tau}_{\mathcal{S}_{k^*}},
    \end{equation} balances improved precision (from borrowing additional controls) against potential bias (from including incompatible controls). 
    \end{itemize}

{\bf Implementation.}  We can implment the above procedures with the \texttt{find\_optimal\_k()} and \texttt{estimate\_selected()} functions. 
  
\begin{lstlisting}[language=R]
result_selected <- find_optimal_k(dat_rct = data_rct, dat_ec = data_ec, 
               influences = influences, reference_value = ATE_true,
               k_vector = seq(5, 120, by = 5))
print(result_selected$mse_optimal) 
  top_k   estimate       bias     variance       mse
   55   -0.9845352   0.01546484   0.02738979 0.02762895
\end{lstlisting}
                      
\begin{figure}[h]
\vspace{-12pt}
    \centering
    \includegraphics[width=0.7\textwidth]{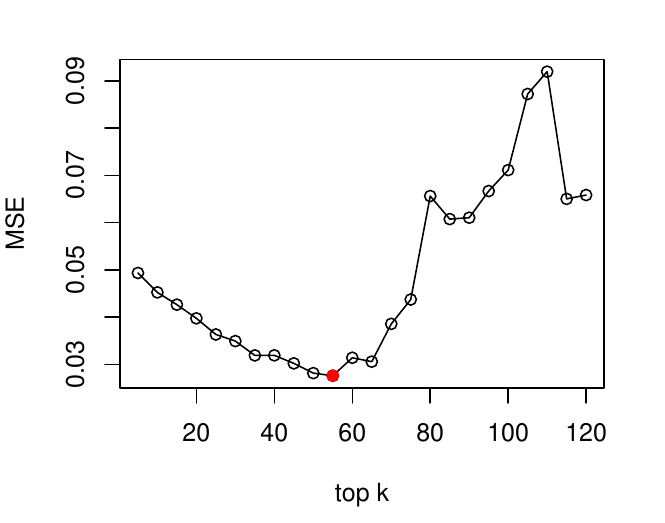}
    \vspace{-16pt}
    \caption{The MSE for varying levels of top-$k$ ECs.} 
    \label{fig7} 
    \vspace{-16pt}
\end{figure}

We present the MSE for different values of $k$ (top-$k$), shown in Figure \ref{fig7}, where the optimal $k$ is 55 (marked in red).   
The corresponding code is given below. 
\begin{lstlisting}[language=R]
mse_k <- result_selected$mse_k
par(mfrow = c(1, 1))
plot(mse_k$top_k, mse_k$mse, type = 'o', xlab = 'top k', ylab = 'MSE')
points(result_selected$mse_optimal$top_k, 
       result_selected$mse_optimal$mse,  col = 'red', pch = 19)
\end{lstlisting}

 In \texttt{find\_optimal\_k()}, we estimate the nuisance parameters by default using linear regression for $m_{\mathcal{S}_k, 0}(x)$ and $m_{\mathcal{S}_k, 1}(x)$, and logistic regression for \( e_{\mathcal{S}_k}(x) \). To incorporate more flexible methods for nuisance parameter estimation, we recommend using the \texttt{estimate\_selected()} function, which supports explicitly modeling nuisance parameters.  
We provide the code in Section C of Supplementary Material, where kernel-based regularized least squares (a machine learning method) is used to estimate $m_{\mathcal{S}_k, 0}(x)$ and $m_{\mathcal{S}_k, 1}(x)$ implemented by \texttt{R} package {\bf KRLS}~\citep{Ferwerda2017KRLS}. The corresponding numerical results are similar to those shown in Figure~\ref{fig7} and are therefore omitted for brevity.

\subsection{Step 5: Comparison of Estimators and Sensitivity Analysis} \label{sec4-6}

Transparent diagnostics and comprehensive reporting are essential for the responsible application of the AIB method. They serve two purposes: (1) to communicate the rationale and uncertainty of the final estimate to collaborators and regulators, and (2) to assess the robustness of the conclusions. This section outlines the key outputs and sensitivity analyses. 

\subsubsection{Comparison of Estimators}   \label{sec4-6-1}

We suggest comparing $\hat \tau_{\text{aib}}$ defined in \eqref{eq6}, with the following benchmark estimators. 

\begin{itemize}
    \item \textbf{RCT-only estimators ($\hat \tau_{\text{dirct}} $ and $\hat \tau_{\textup{aipw}}$)}. It contains the direct contrast of treated outcome and control outcomes, 
    			\begin{equation}    \hat \tau_{\text{dirct}} =   \frac{1}{ n_{1t} } \sum_{\{i: A_i =1\}} Y_i - \frac{1}{ n_{1c} } \sum_{\{i: A_i =0\}} Y_i,           \end{equation}
 where $n_{1t}$ and $n_{1c}$ are the numbers of treated and control units in the RCT data, respectively. It is also recommended to report $\hat \tau_{\textup{aipw}}$, as defined in \eqref{eq-AIPW}. 
    
    \item \textbf{Full-borrowing estimator ($\hat{\tau}_{\text{full}}$)}.  
The treatment effect estimate obtained by naively pooling all ECs is given by
\begin{equation}
\hat{\tau}_{\text{full}} = \hat{\tau}_{\mathcal{S}_{n_0}},
\end{equation}
that is, it corresponds to the estimator $\hat{\tau}_{\mathcal{S}_k}$ with $k = n_0$, the total number of ECs.   
\end{itemize}

We implement the RCT-only estimators using \texttt{estimate\_rct()} function and the full-borrowing estimator using  \texttt{estimate\_selected()} function.   
\begin{lstlisting}[language=R]
n_rct <- nrow(data_rct)
n_ec <- nrow(data_ec)
p <- ncol(data_rct) - 2  
# RCT-only estimators
X <- data_rct[, 1:p]
A <- data_rct$A
Y <- data_rct$Y
result_rct <- estimate_rct(X, A, Y)  
# Full-borrowing estimator 
data_combine <- rbind(data_rct, data_ec)
X <- data_combine[, 1:p]
A <- data_combine$A
Y <- data_combine$Y
result_full <- estimate_selected(X, A, Y, reference_value = ATE_true) 
# print the estimators
result <- data.frame(
  estimate = c(result_rct$estimate, result_full$estimate,  
               result_selected$mse_optimal$estimate),
  bias = c(result_rct$estimate, result_full$estimate,  
           result_selected$mse_optimal$estimate) - ATE_true, 
  std = c(result_rct$se, result_full$se, 
          sqrt(result_selected$mse_optimal$variance)),
  row.names = c('direct', 'aipw', 'full', 'aib'))
result$mse <- result$bias^2 + result$std^2 
print(result)
              estimate        bias       std        mse
direct      -0.9815725 0.018427529 0.3922757 0.15421978
aipw        -0.9989774 0.001022596 0.2323241 0.05397554
full        -0.9294999 0.070500133 0.1828357 0.03839918
aib         -0.9845352 0.015464844 0.1654986 0.02762895
\end{lstlisting}
From the results (the last five lines above), $\hat \tau_{\text{aib}} =  \hat \tau_{\mathcal{S}_{k^*}}$ ($k^* = 55$) achieves better performance than the other estimators in terms of lower MSE. 

%



\subsubsection{Sensitivity Analyses}  

To assess the robustness of the results from the AIB method, we recommend conducting and reporting the following sensitivity analyses. 

 \textbf{Varying \(k\) around the optimum:} Report treatment effect estimates and confidence intervals for \(k^* \pm \delta\), where \(\delta\) is a small integer (e.g., 10\% of \(n_{0}\)). 
This helps assess how sensitive the conclusions are to the specific choice of the borrowing cutoff. 
From Figure \ref{fig7},  $k^* = 55$; we therefore examine $k$ in the range from 40 to 70, with increments of 1. 

\begin{lstlisting}[language=R]
k_vector <- seq(40, 70, by = 1) 
result_sen <- find_optimal_k(dat_rct = data_rct, dat_ec = data_ec, 
       influences = influences, reference_value = ATE_true, 
       k_vector = k_vector)
print(result_sen$mse_k)
\end{lstlisting}
 \begin{figure}[h]
\vspace{-16pt}
    \centering
    \includegraphics[width=0.7\textwidth]{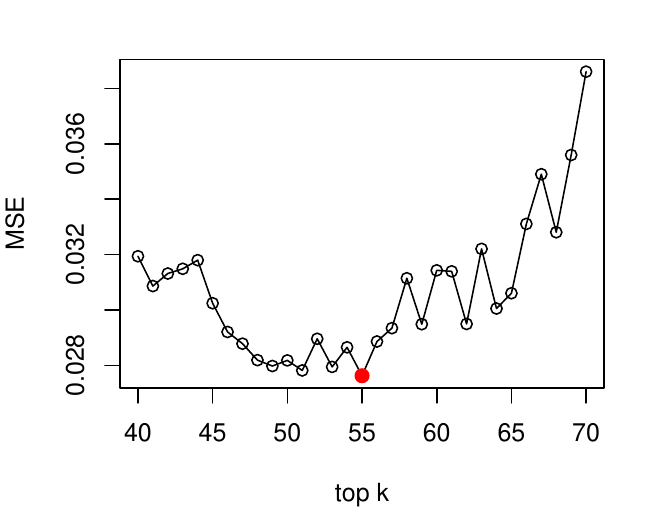}
    \vspace{-16pt}
    \caption{The MSE for varying levels of top-$k$ ECs.} 
    \label{fig8} 
\end{figure}
From Figure \ref{fig8}, the MSE of $\hat \tau_{\mathcal{S}_k}$ for $k$ around $k^* = 55$ is relatively stable, varying from $0.028$ to $0.036$. 
We also note that the MSE curve fluctuates slightly around $k^*$; this is reasonable and may arise because, when additional ECs around $k^*$ are included, they may exert effects in different directions on $\hat{\mu}_0(x)$ (e.g., in a linear model, one may slightly increase the slope while another may slightly decrease it).

\textbf{Alternative nuisance parameter model specifications:}  
Re-run the entire workflow using different models for \({\mu}_0(X)\) in Step 1 (e.g., logistic vs. probit regression for binary outcomes, adding interaction terms). We omit them for compactness.  

\textbf{Additional numerical studies.}  
In Section D of the Supplementary Material, we present additional numerical studies demonstrating the practical implementation of the AIB framework under different data-generating mechanisms. Specifically, we consider two scenarios: one with a binary outcome and another with a continuous outcome generated from a nonlinear mechanism. 
\section{Outcome Calibration for Improving Compatibility}
\label{sec5}

The AIB framework described in Section 4 identifies comparable EC units by selecting those whose covariate-outcome patterns most closely resemble those of RCT controls. This approach works well when a reasonable fraction of ECs are individually compatible with the trial population. However, a different and practically important scenario arises when \emph{most} ECs are systematically different from RCT controls, e.g., because the EC data come from an earlier period when the standard of care was less effective, from a different geographic region with distinct clinical practices, or from a healthcare system with different treatment protocols. In such settings, the influence-based selection in Step 2 may retain only a small fraction of ECs, limiting the efficiency gains that motivated external borrowing in the first place.
This brings us to the third key methodological question: 
\begin{tcolorbox}
\textbf{Q3}: When only a small fraction of ECs are deemed comparable, can we expand the usable pool through outcome calibration?  
\end{tcolorbox}
This section provides a detailed description of Step 0 in Section \ref{sec4-1}, to answer  Q3.

\subsection{Calibrating Differences in the Covariate-Outcome  Pattern} 
The calibration procedure introduces a bias function $b(x)$ to capture the systematic difference in control outcomes between ECs and RCT controls as a function of covariates~\citep{Yang-etal2025-Bernoulli}. Specifically, the observed outcome $Y_i$ of each EC unit is adjusted as
\begin{equation*} 
\tilde{Y}_i = Y_i - b(X_i), \quad \text{for all } i \in \mathcal{D}_{\text{EC}}. 
\end{equation*}
The bias function $b(x)$ is defined such that, after adjustment, the conditional mean of the calibrated EC outcome matches that of the RCT control outcome for units with the same covariates:
$\mathbb{E}_{\P}[\tilde{Y} \mid X=x, A=0] = \mathbb{E}_{\mathbb{Q}}[Y \mid X=x, A=0]$,
ensuring that the covariate-outcome patterns between ECs and RCT controls are aligned. 

Let $\tilde \P := \P \cup \mathbb{Q}$ be the combined population of $\P$ and $\mathbb{Q}$, and let $R$ denote the data source indicator, where $R=1$ indicates an RCT unit and $R=0$ indicates an EC unit. 
 Following R-learner framework~\citep{Nie-Wager2021, Wu-Shu-Rlearner}, we  estimate $b(x)$ by minimizing   
	        \[       \sum_{ \{i \in \mathcal{D}_{\text{RCT}}\cup  \mathcal{D}_{\text{EC}}:  A_i = 0\} } \big (  Y_i - \hat m(X_i) -  (\hat \pi_0(X_i) - R_i ) b(X_i) \big )^2      \]
where $\pi_0(x) = \mathbb{E}_{\tilde \P}[R \mid X=x, A=0]$ denotes the sampling  score among the controls, 
$m(x) = \mathbb{E}_{\tilde \P}[Y \mid X=x, A=0]$ denote the outcome regression function for all controls,  and $\hat m(x)$ and $\hat \pi_0(x)$ are their estimates.  
If $b(x)$ is estimated, we subtract the corresponding estimated bias from each EC's observed outcome to obtain an adjusted EC dataset. We then apply the AIB method (Section~\ref{sec4}) using these \emph{calibrated} ECs. {\bf The final estimate from the calibrated AIB (CAIB) method is denoted by  $\hat \tau_{\text{caib}}$.} 

Intuitively, calibration corrects the \emph{average} discrepancy, but some individual ECs may still be noisy or atypical. The subsequent influence-based filtering step (Step 4 of  the AIB framework) ensures that we borrow only those calibrated units that are both well adjusted and individually compatible.  


\subsection{Implementation}
For implementation, the bias function \( b(x) \) can be estimated using the \texttt{rlearner\_lm()} and \texttt{rlearner\_krls()} functions. The former specifies linear models (including logistic regression) for \( m(x) \), \( \pi_0(x) \), and \( b(x) \), whereas the latter employs kernel-based regularized least squares, a nonparametric approach that offers greater modeling flexibility. We use \texttt{rlearner\_lm()} as an example.  

\begin{lstlisting}[language=R]
data_combine$R <- c( rep(1, n_rct), rep(0, n_ec))
data_combine_control <- data_combine[data_combine$A ==0, ]
X <- data_combine_control[, 1:p]
Y <- data_combine_control$Y
R <- data_combine_control$R  
# estimating b(x) with linear model
rlm_fit <- rlearner_lm(X, R, Y)
X_ec <- data_ec[, 1:p]
bias_est <- predict(rlm_fit, X_ec)
data_ec_calibrated <- data_ec
data_ec_calibrated$Y <- data_ec$Y - bias_est
\end{lstlisting}

 \begin{figure}[h]
\vspace{-8pt}
    \centering
    \includegraphics[width=1\textwidth]{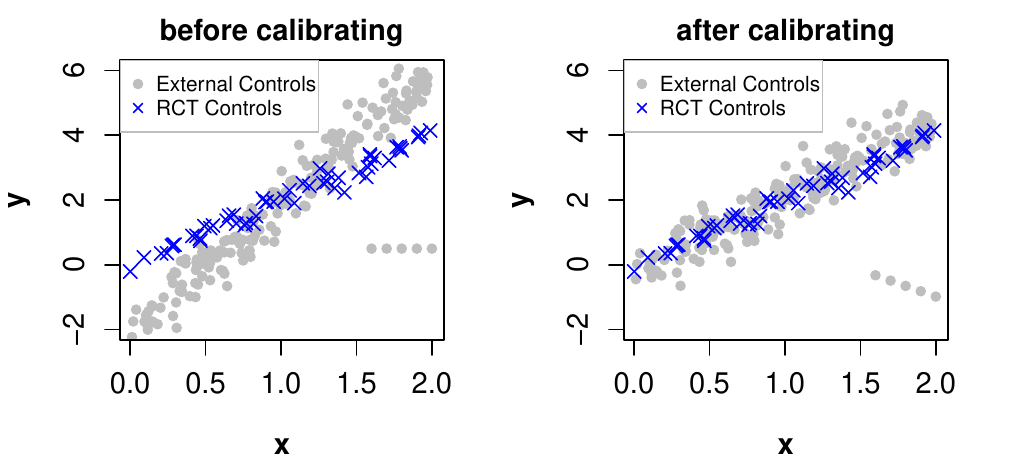}
    \vspace{-46pt}
    \caption{Comparison of scatter plots before and after outcome calibration.} 
    \label{fig9} 
\end{figure}

Figure~\ref{fig9} illustrates the differences in scatter plots for RCT controls and ECs, with and without outcome calibration. The results indicate that outcome calibration can well adjust systematic differences between the two groups.    
After calibration, we apply the AIB method to the calibrated ECs, and the corresponding results are reported in Table \ref{table1}. The results indicate that outcome calibration  increases the number of comparable samples in ECs and yields an estimator with a smaller MSE. 
 
\begin{table}[h]  
\caption{Comparison of various estimators}
\centering
\begin{tabular}{lccccc}
  \hline
 & Estimate & Bias & SD & MSE &  $k^*$ \\ 
  \hline
Direct ($\hat \tau_{\text{direct}}$) & -0.982 & 0.018 & 0.392 & 0.154 & 0   \\ 
  AIPW ($\hat{\tau}_{\text{aipw}}$)  & -0.999 & 0.001 & 0.232 & 0.054 & 0  \\ 
  Full ($\hat{\tau}_{\text{full}}$) & -0.929 & 0.071 & 0.183 & 0.038 & 200 \\ 
  Selected 
($\hat{\tau}_{\text{aib}}$) &  -0.985 & 0.015 & 0.165 & 0.028 & 55 \\ 
  Calibrated \& Selected ($\hat{\tau}_{\text{caib}}$)  & -1.011 & -0.011 & 0.143 & 0.020 & 185 \\ 
   \hline  
\end{tabular}  \medskip 
\begin{flushleft}  
Note: 
Bias, SD,  MSE denote the bias, standard deviation, mean squared error of the estimators, respectively. 
\end{flushleft} \label{table1}
\end{table}

{\bf Practical recommendation}. We recommend running the AIB workflow both with and without outcome calibration and reporting both sets of results. If the two approaches yield similar conclusions, calibration provides an additional efficiency gain at modest computational cost, and the results can be presented together as mutually reinforcing evidence. If the two approaches diverge meaningfully, the discrepancy itself is informative and warrants further investigation into the nature and extent of the systematic differences between the EC and RCT populations.

\section{Application to the NSW--PSID Study}  \label{sec6}
We further demonstrate the AIB framework using real-world datasets.

{\bf Data Description}. 
We conduct our empirical study using the National Supported Work (NSW) dataset~\citep{lalonde1986evaluating} as the RCT sample and the Population Survey of Income Dynamics (PSID) dataset~\citep{dehejia2002propensity} as the EC sample. The NSW program was designed to evaluate whether structured job training and supported work experience could improve labor market outcomes for economically disadvantaged individuals. The NSW dataset includes 445 randomized participants, comprising 185 treated units and 260 control units. The PSID dataset includes 128 control units. The treatment $A$ equals 1 for units enrolled in the NSW job training program and 0 otherwise. The outcome $Y$ is 1978 earnings (RE78). The baseline covariates $X$ include age, education, race, Hispanic ethnicity, marital status, high school degree attainment, and pre-treatment earnings in 1974 and 1975 (RE74 and RE75). All earnings variables are rescaled by a factor of 1,000 for analysis.


%

{\bf Setup.}  In this application, we take the NSW-only (RCT-only) AIPW estimator as the reference and assess whether adaptively incorporating ECs improves efficiency relative to this benchmark.  
For the AIB and  comparing Direct and AIPW methods, we use logistic regression to estimate $e_{\mathcal{S}_k}(x)$ and $e_1(x)$, and linear model to estimate $m_{\mathcal{S}_k, 0}(x)$, $m_{\mathcal{S}_k, 1}(x)$, $\mu_0(x)$, and $\mu_1(x)$.    
  For the outcome calibration in the CAIB method, we use logistic regression to estimate $\pi_0(x)$ and linear model to estimate $m(x)$ and $b(x)$. 
We use the Akaike information criterion (AIC) to select covariates for the two linear working models of $\mu_1(x)$ and $\mu_0(x)$, and take the union of the selected covariates sets for subsequent analysis.

\begin{table}[htbp]
	\centering
	\caption{Results for the NSW--PSID study.}
	\label{tab:nsw_psid_results}
	\begin{tabular}{lccccc}
		\toprule
		Estimator & Estimate & Bias & SD & MSE & $k^*$ \\
		\midrule
		Direct ($\hat{\tau}_{\mathrm{direct}}$) & 1.79434 & 0.06760 & 0.86029 & 0.74467 & 0 \\
		AIPW ($\hat{\tau}_{\mathrm{aipw}}$) & 1.72674 & 0.00000 & 0.64147 & 0.41148 & 0 \\
		Full ($\hat{\tau}_{\mathrm{full}}$) & 1.81878 & 0.09204 & 0.63440 & 0.41093 & 128 \\
		Selected ($\hat{\tau}_{\mathrm{aib}}$) & 1.69324 & -0.03351 & 0.63662 & 0.40640 & 10 \\
		Calibrated \& Selected ($\hat{\tau}_{\mathrm{caib}}$) & 1.76662 & 0.03988 & 0.60487 & {\bf 0.36746} & 50 \\
		\bottomrule
	\end{tabular}
	\medskip 
	\begin{flushleft}
	\end{flushleft}
\end{table}


{\bf Results}.  
The numerical results are reported in Table~\ref{tab:nsw_psid_results}.
From it, we have the following observations. (a) The two NSW-only estimators ($\hat{\tau}_{\mathrm{direct}}$ and $\hat{\tau}_{\mathrm{aipw}}$) both suggest a positive effect of the job training program on earnings, with the AIPW estimator $\hat{\tau}_{\mathrm{aipw}}$ exhibiting a smaller SD than the direct estimator $\hat{\tau}_{\mathrm{direct}}$.  
(b) Full borrowing ($\hat{\tau}_{\mathrm{full}}$) incorporates all 128 PSID controls  and slightly reduces the SD relative to the AIPW estimator. However, it also shifts the point estimate further away from the AIPW reference, resulting in only a marginal reduction in MSE.  
(c) The AIB method ($\hat{\tau}_{\mathrm{aib}}$) selects only 10 PSID controls. This implicitly indicates that the regression functions for the control outcome differ substantially between the NSW controls and the PSID controls, leading the AIB method to select only a small subset of PSID controls. 
In addition, compared with full borrowing, the AIB method yields an estimate closer to the AIPW reference and a smaller MSE, demonstrating the efficiency of the AIB method.  
(d) The CAIB method ($\hat{\tau}_{\mathrm{caib}}$)  selects 50 PSID controls and achieves the smallest MSE among all methods. This suggests that outcome calibration effectively increases the number of comparable ECs and leads to further efficiency improvements.

\section{Conclusion}  \label{sec7}

This tutorial provides a comprehensive guide to the AIB framework, a novel approach for augmenting RCTs with EC data. 
The AIB framework rests on a simple but powerful idea: rather than making a binary decision about whether to borrow from the entire EC dataset, it assesses compatibility at the individual patient level. By translating the abstract notion of exchangeability into a concrete, computable influence score for each EC unit, the framework enables a nuanced, data-driven borrowing decision that adapts to the heterogeneity inherent in real-world EC datasets. The key insights from this tutorial can be summarized as follows:

\quad  $\bullet$ Individual-level assessment is more informative than population-level decisions. Real-world EC datasets are rarely uniformly compatible or incompatible with RCT controls. Some ECs closely resemble RCT controls; others do not. Treating all ECs identically,  
 fails to exploit this heterogeneity productively. The influence score provides a principled, patient-level metric that captures this variation directly.
 
 \quad  $\bullet$  MSE minimization provides a disciplined bias-variance trade-off. The nested subset construction and MSE-based selection in Steps 3 and 4 operationalize the bias-variance trade-off in a transparent and data-driven way. The resulting MSE curve provides a visual and quantitative summary of how borrowing more ECs progressively reduces variance but may increase bias, and the optimal $k$ identifies the point at which these competing forces are best balanced.
  
   \quad  $\bullet$ Outcome calibration and influence-based selection are complementary. Calibration addresses systematic, population-level differences between ECs and RCT controls, expanding the pool of units that are individually compatible after adjustment. Influence-based selection then guards against residual individual-level incompatibility among the calibrated units. Together, they provide a two-stage defense against bias that is both flexible and robust.
   
  \quad  $\bullet$  The framework is assumption-lean and broadly applicable. Unlike many borrowing methods that impose parametric models on the EC outcome distribution, the AIB framework requires no outcome model for the EC population. Influence scores are computed using only the RCT control outcome model, making the approach robust to misspecification of the EC distribution and broadly applicable across diverse clinical settings and outcome types.

Nevertheless, practical implementation may still face challenges. One common issue is over-borrowing driven by variance reduction pressure. The MSE-minimization criterion balances bias and variance, but in very small RCTs, the variance component can dominate, pushing the optimal $k$ toward borrowing more samples even when they introduce bias. This may lead to an undesirable trade-off that favors precision over validity. 

To address this, one should carefully inspect the MSE-$k$ curve; if it is relatively flat around the minimum, a slightly smaller $k$ may be preferable as a more conservative choice, accepting a modest increase in MSE to reduce potential bias. It is also important to perform a clinical plausibility check: if borrowing many ECs leads to estimates that deviate substantially from the RCT-only result, the discrepancy should be discussed with domain experts. Finally, the RCT-only estimate should serve as an anchor, as regulatory considerations typically prioritize internal validity. The adaptive estimate should therefore be viewed as complementary evidence rather than a replacement unless strong comparability is established.

\appendix


\section{Disclosure statement}\label{disclosure-statement}
The authors declare no conflicts of interest.

\section{Data Availability Statement}\label{data-availability-statement}

The NSW and PSID data used in the application are available at \url{https://users.nber.org/~rdehejia/nswdata2.html}.

\phantomsection\label{supplementary-material}
\bigskip

\begin{center}

{\large\bf SUPPLEMENTARY MATERIAL}

\end{center}

%

\section{Flexible Estimation of Outcome Regression Functions}

The default nuisance models in \texttt{find\_optimal\_k()} use linear regression for $m_{\mathcal{S}_k, 0}(x)$ and $m_{\mathcal{S}_k, 1}(x)$, and logistic regression for \( e_{\mathcal{S}_k}(x) \). To incorporate more flexible methods for nuisance parameter estimation, we recommend using the \texttt{estimate\_selected()} function, which supports explicitly modeling nuisance parameters.  
We provide the illustrative codes, where kernel-based regularized least squares (a machine learning method) is used to estimate $m_{\mathcal{S}_k, 0}(x)$ and $m_{\mathcal{S}_k, 1}(x)$ implemented by \texttt{R} package \texttt{KRLS}~\citep{Ferwerda2017KRLS}.  
   
 \begin{lstlisting}[language=R]               
library(KRLS)
k_vector <- seq(5, 120, by = 5)
mse_k <- data.frame(top_k = k_vector,  # used for save results
                         mse = rep(NA, length(k_vector)),
                         bias = rep(NA, length(k_vector)),
                         variance = rep(NA, length(k_vector)))
n_rct <- nrow(data_rct)
p <- ncol(data_rct) - 2
sorted_indices <- order(influences, decreasing = FALSE)
for(i in 1:length(k_vector)){
  top_k <- k_vector[i]
  # selected data
  selected_indices <- sorted_indices[1:top_k]
  dat_selected <- rbind(data_rct, data_ec[selected_indices, ])
  X <- dat_selected[, 1:p, drop = F]
  A <- dat_selected$A
  Y <- dat_selected$Y
  # estimating nuisance parameters
  # estimating e1(x) using logistic regression
  data_ps = data.frame(A = A, X = X)
  model_ps = glm(A ~ ., data = data_ps, family = binomial())
  ps_hat = predict(model_ps, type = "response")
  ps_hat = unname(ps_hat)
  # estimating m1(x) and m0(x) using kernel-based regularized least squares
  model_m1 <- krls(X = X[A==1, ], y = Y[A==1])
  m1_hat <- predict(model_m1, newdata = X)$fit 
  model_m0 <- krls(X = X[A==0, ], y = Y[A==0])
  m0_hat <- predict(model_m0, newdata = X)$fit 
  # obtain the estimator of ATE
  mod <- estimate_selected(X, A, Y, reference_value = ATE_true,
                 ps_hat = ps_hat, mu0_hat = m0_hat, mu1_hat = m1_hat)
  mse_k$mse[i] <- mod$mse
  mse_k$bias[i] <- (mod$estimate - ATE_true)
  mse_k$variance[i] <- mod$se^2
}
# the optimal one
mse_optimal <- mse_k[which.min(mse_k$mse), ] 
plot(mse_k$top_k, mse_k$mse, type = 'o', xlab = 'top k', ylab = 'MSE')
points(mse_optimal$top_k, mse_optimal$mse, col = 'red', pch = 19)
\end{lstlisting}

\section{Additional Simulation}

In this section, we present additional numerical studies to demonstrate the practical implementation of the adaptive influence-based borrowing framework under complex data-generating mechanisms. Specifically, we consider two distinct scenarios: one with a binary outcome, and another with a continuous outcome generated from a nonlinear mechanism with two-dimensional covariates.

\subsection{Numerical Study 1: Binary Outcomes}

In the first scenario, we consider a single covariate $X$ and a binary outcome $Y$. The RCT data ($n_1 = 100$) is generated using a standard logistic outcome model. The external control (EC) data ($n_0 = 400$) is generated with a quadratic bias term and incorporates deterministic outliers at the right tail of the covariate distribution. 

Specifically, the covariate $X$ is generated from a Uniform$(0, 2)$ distribution, with 20 outlier samples intentionally placed in the interval $[1.8, 2.0]$. The outcome generating mechanisms are defined as follows:
	
		RCT Data ($\mathcal{D}_{\text{RCT}}$): The treatment assignment is randomized with $\P(A=1) = 0.5$. The potential outcomes are generated from a logistic model
		$\P(Y(a)=1 \mid X) = \text{expit}(X - 1 + 2a), \quad \text{for } a \in \{0, 1\}$,
	where $\text{expit}(v) = 1 / \{1 + \exp(-v)\}$. The true average treatment effect is $\tau \approx 0.368$.
	
		EC Data ($\mathcal{D}_{\text{EC}}$): All participants are in the control arm ($A=0$). The outcome incorporates a quadratic bias
		$\mathbb{Q}(Y=1 \mid X, A=0) = \text{expit}(X - 1 + 2.5(X - 1)^2)$.
	Additionally, to introduce severe non-comparability, the 20 right-tail outlier samples are deterministically assigned $Y = 1$.

The following code chunk demonstrates the core evaluation workflow using the \texttt{InfluenceBorrowing} package, assuming the data has been generated and loaded as \texttt{Dat1}.

\begin{lstlisting}[language=R]
	# =================== Test Mechanism 1 =================== 
	cat("\n--- Mechanism 1: 1D Covariate, Logistic Model ---\n")
	set.seed(42)
	x_large <- runif(5000000, 0, 2)
	ATE_true_1 <- mean(plogis(x_large + 1) - plogis(x_large - 1))
	
	Dat1 <- gen_dat_mech1(n_rct = 100, n_ec = 400, seed = 2026)
	p_1 <- 1
	
	# 1. Baseline Estimates
	result_rct1 <- estimate_rct(Dat1$data_rct[, 1:p_1, drop=FALSE], Dat1$data_rct$A, Dat1$data_rct$Y, outcome_family = binomial())
	result_full1 <- estimate_selected(
		X = Dat1$dat_combine[, 1:p_1, drop=FALSE], 
		A = Dat1$dat_combine$A, 
		Y = Dat1$dat_combine$Y, 
		reference_value = result_rct1$estimate["aipw"], 
		outcome_family = binomial()
	)
	
	# 2. AIB Method
	data_rct_control1 <- Dat1$data_rct[Dat1$data_rct$A == 0, ]
	
	# fit outcome model on RCT controls (step 1)
	model1 <- glm(Y ~ X, data = data_rct_control1, family = binomial(link = "logit"))
	
	# calculate influence score and find optimal subset based on MSE (step 2,3,4)
	test_data1 <- data.frame(X = Dat1$data_ec$X, Y = Dat1$data_ec$Y)
	influences1 <- compute_influences(model1, testdata = test_data1)
	
	optimal_res1 <- find_optimal_k(
		dat_rct = Dat1$data_rct,
		dat_ec = Dat1$data_ec,
		influences = influences1,
		reference_value = result_rct1$estimate["aipw"],
		trim = 0.01, 
		k_vector = seq(0, nrow(Dat1$data_ec), by = 10), 
		outcome_family = binomial()
	)
	
	# 3. ACIB Method
	data_combine_control1 <- Dat1$dat_combine[Dat1$dat_combine$A == 0, ]
	X_comb_ctrl1 <- data_combine_control1[, 1:p_1, drop=FALSE]
	Y_comb_ctrl1 <- data_combine_control1$Y
	R_comb_ctrl1 <- data_combine_control1$R
	
	# fit outcome calibration model using rlearner_lm and calibrate ECs (step 0)
	rlm_fit1 <- rlearner_lm(X_comb_ctrl1, R_comb_ctrl1, Y_comb_ctrl1)
	X_ec1 <- Dat1$data_ec[, 1:p_1, drop=FALSE]
	bias_est1 <- predict(rlm_fit1, X_ec1)
	
	data_ec_calibrated1 <- Dat1$data_ec
	data_ec_calibrated1$Y <- Dat1$data_ec$Y - bias_est1
	
	# calculate new influence scores and find optimal subset on calibrated data (step 2,3,4)
	test_data_cal1 <- data.frame(X = data_ec_calibrated1$X, Y = data_ec_calibrated1$Y)
	influences_cal1 <- compute_influences(model1, testdata = test_data_cal1)
	
	optimal_res_cal1 <- find_optimal_k(
		dat_rct = Dat1$data_rct,
		dat_ec = data_ec_calibrated1,
		influences = influences_cal1,
		reference_value = result_rct1$estimate["aipw"], 
		trim = 0.01, 
		k_vector = seq(0, nrow(Dat1$data_ec), by = 10)
	)
	
	# 4. Print results for Mechanism 1
	res_df1 <- create_res_df(result_rct1, result_full1, optimal_res1, optimal_res_cal1, ATE_true_1, nrow(Dat1$data_ec))
	print(round(res_df1, 3))
\end{lstlisting}

{
Table~\ref{tab:study1} summarizes the performance of different estimators. Na\"ive pooling of all external controls (\texttt{Full}) introduces substantial bias ($-0.122$). In contrast, the adaptive influence-based estimator ($\hat{\tau}_{\mathcal{S}_{k^*}}$) limits the inclusion of non-comparable samples. It achieves a bias ($0.068$) comparable to the RCT-only AIPW estimator ($\hat{\tau}_{\text{aipw}}$), while simultaneously improving precision by reducing the standard deviation from $0.084$ to $0.061$. The calibrated adaptive estimator further improves the bias-variance trade-off, with bias $0.056$, standard deviation $0.057$, and the smallest MSE among all estimators.
}

\begin{table}[ht]
	\centering
	\caption{Comparison of estimators for Mechanism 1.}
	\label{tab:study1}
	\begin{tabular}{lccccc}
		\toprule
		\textbf{Estimator} & \textbf{Estimate} & \textbf{Bias} & \textbf{SD} & \textbf{MSE} & \textbf{$k^*$} \\
		\midrule
		Direct ($\hat{\tau}_{\text{direct}}$) & 0.435 & 0.067 & 0.160 & 0.030 & 0 \\
		AIPW ($\hat{\tau}_{\text{aipw}}$) & 0.435 & 0.067 & 0.084 & 0.012 & 0 \\
		Full ($\hat{\tau}_{\text{full}}$) & 0.246 & -0.122 & 0.057 & 0.018 & 400 \\
		Selected ($\hat{\tau}_{\mathcal{S}_{k^*}}$) & 0.435 & 0.068 & 0.061 & 0.004 & 220 \\
		Calibrated \& Selected & 0.424 & 0.056 & 0.057 & 0.003 & 400 \\
		\bottomrule
	\end{tabular}
\end{table}

\subsection{Numerical Study 2: Non-linear Model}

In the second scenario, we consider a continuous outcome setting with a two-dimensional covariate space, $X = (X_1, X_2)$. The true average treatment effect (ATE) in the RCT is constantly 3. The external control data is generated with a non-linear cubic term and includes outliers located at the corner of the covariate space.

The covariates $(X_1, X_2)$ are generated from independent Uniform$(0, 2)$ distributions, with 20 outlier samples deterministically placed in the region $[1.8, 2.0] \times [1.8, 2.0]$. The outcome generating mechanisms are specified as follows:
		RCT Data ($\mathcal{D}_{\text{RCT}}$): The sample size is $n_1 = 100$, and treatment is randomized with $\P(A=1) = 0.5$. The potential outcomes are generated as
		$Y(0) = 2X_1 + 2X_2 + \varepsilon, 
		Y(1) = Y(0) + 3$.
	where $\varepsilon \sim \mathcal{N}(0, 0.5^2)$. The true ATE is exactly $\tau = 3$. 
	
		EC Data ($\mathcal{D}_{\text{EC}}$): The sample size is $n_0 = 400$, and all participants are controls ($A=0$). The outcome incorporates a non-linear cubic term
		$Y = -2 + 4X_1 + 2X_2 + 2(X_1 - 1)^3 + \varepsilon$. 
	Additionally, to emulate structural discrepancy, the 20 corner outlier samples are deterministically assigned an extreme value of $Y = -5$.

The following code illustrates the method implementation for this 2D continuous outcome scenario:

\begin{lstlisting}[language=R]
	# =================== Test Mechanism 2 =================== 
	cat("\n- Mechanism 2: 2D Covariates, Continuous Non-linear Model -\n")
	ATE_true_2 <- 3 # the true value of ATE in RCT
	
	Dat2 <- gen_dat_mech2(n_rct = 100, n_ec = 400, seed = 2026)
	p_2 <- 2
	
	# 1. Baseline Estimates
	result_rct2 <- estimate_rct(Dat2$data_rct[, 1:p_2, drop=FALSE], Dat2$data_rct$A, Dat2$data_rct$Y)
	result_full2 <- estimate_selected(
		X = Dat2$dat_combine[, 1:p_2, drop=FALSE], 
		A = Dat2$dat_combine$A, 
		Y = Dat2$dat_combine$Y, 
		reference_value = result_rct2$estimate["aipw"]
	)
	
	# 2. AIB Method (Influence-based without calibration)
	data_rct_control2 <- Dat2$data_rct[Dat2$data_rct$A == 0, ]
	
	# fit outcome model on RCT controls (step 1)
	model2 <- glm(Y ~ X1 + X2, data = data_rct_control2, family = gaussian(link = "identity"))
	
	# calculate influence score and find optimal subset based on MSE (step 2,3,4)
	test_data2 <- data.frame(X1 = Dat2$data_ec$X1, X2 = Dat2$data_ec$X2, Y = Dat2$data_ec$Y)
	influences2 <- compute_influences(model2, testdata = test_data2)  
	
	optimal_res2 <- find_optimal_k(
		dat_rct = Dat2$data_rct,
		dat_ec = Dat2$data_ec,
		influences = influences2,
		reference_value = result_rct2$estimate["aipw"], 
		trim = 0.01, 
		k_vector = seq(0, nrow(Dat2$data_ec), by = 10)
	)
	
	# 3. ACIB Method
	data_combine_control2 <- Dat2$dat_combine[Dat2$dat_combine$A == 0, ]
	X_comb_ctrl2 <- data_combine_control2[, 1:p_2, drop=FALSE]
	Y_comb_ctrl2 <- data_combine_control2$Y
	R_comb_ctrl2 <- data_combine_control2$R
	
	# fit outcome calibration model using rlearner_lm and calibrate ECs (step 0)
	rlm_fit2 <- rlearner_lm(X_comb_ctrl2, R_comb_ctrl2, Y_comb_ctrl2)
	X_ec2 <- Dat2$data_ec[, 1:p_2, drop=FALSE]
	bias_est2 <- predict(rlm_fit2, X_ec2)
	
	data_ec_calibrated2 <- Dat2$data_ec
	data_ec_calibrated2$Y <- Dat2$data_ec$Y - bias_est2
	
	# calculate new influence scores and find optimal subset on calibrated data (step 2,3,4)
	test_data_cal2 <- data.frame(X1 = data_ec_calibrated2$X1, X2 = data_ec_calibrated2$X2, Y = data_ec_calibrated2$Y)
	influences_cal2 <- compute_influences(model2, testdata = test_data_cal2)
	
	optimal_res_cal2 <- find_optimal_k(
		dat_rct = Dat2$data_rct,
		dat_ec = data_ec_calibrated2,
		influences = influences_cal2,
		reference_value = result_rct2$estimate["aipw"], 
		trim = 0.01, 
		k_vector = seq(0, nrow(Dat2$data_ec), by = 10)
	)
	
	# 4. Print results for Mechanism 2
	res_df2 <- create_res_df(result_rct2, result_full2, optimal_res2, optimal_res_cal2, ATE_true_2, nrow(Dat2$data_ec))
	print(round(res_df2, 3))
	
	# 5. Plot mechanism 2
	optimal_k2 <- optimal_res2$mse_optimal$top_k
	selected_idx2 <- order(influences2)[1:optimal_k2]
	
	pdf("suppl_fig1.pdf",
		width = 7.0,
		height = 3.6,
		family = "Helvetica")
	par(mfrow = c(1, 2), mar = c(4, 4, 2, 1), mgp = c(2.2, 0.7, 0), tcl = -0.3, cex = 0.9)
	plot(test_data2$X1, Dat2$data_ec$Y, col = rgb(0.5, 0.5, 0.5, 0.4), pch = 19, cex = 0.7,
	xlab = "X1", ylab = "Y", main = "Mechanism 2: Y vs X1", font.lab = 2)
	points(Dat2$data_rct$X1[Dat2$data_rct$A==0], Dat2$data_rct$Y[Dat2$data_rct$A==0], col = "blue", pch = 4, cex = 1.2, lwd=2)
	if(optimal_k2 > 0) points(test_data2$X1[selected_idx2], Dat2$data_ec$Y[selected_idx2], col = "red", pch = 19, cex = 0.8)
	legend("topleft", legend = c("ECs", "RCT", "Selected ECs"), 
	col = c("gray", "blue", "red"), pch = c(19, 4, 19), cex = 0.8, bg = "white")
	
	plot(test_data2$X2, Dat2$data_ec$Y, col = rgb(0.5, 0.5, 0.5, 0.4), pch = 19, cex = 0.7,
	xlab = "X2", ylab = "Y", main = "Mechanism 2: Y vs X2", font.lab = 2)
	points(Dat2$data_rct$X2[Dat2$data_rct$A==0], Dat2$data_rct$Y[Dat2$data_rct$A==0], col = "blue", pch = 4, cex = 1.2, lwd=2)
	if(optimal_k2 > 0) points(test_data2$X2[selected_idx2], Dat2$data_ec$Y[selected_idx2], col = "red", pch = 19, cex = 0.8)
	legend("topleft", legend = c("ECs", "RCT", "Selected ECs"), 
	col = c("gray", "blue", "red"), pch = c(19, 4, 19), cex = 0.8, bg = "white")
	par(mfrow = c(1, 1))
	dev.off()
	
	# 6. Plot mechanism 2 (Calibrated)
	optimal_k_cal2 <- optimal_res_cal2$mse_optimal$top_k
	selected_idx_cal2 <- order(influences_cal2)[1:optimal_k_cal2]
	
	pdf("suppl_fig2.pdf",
		width = 7.0,
		height = 3.6,
		family = "Helvetica")
	par(mfrow = c(1, 2), mar = c(4, 4, 2, 1), mgp = c(2.2, 0.7, 0), tcl = -0.3, cex = 0.9)
	
	plot(test_data_cal2$X1, data_ec_calibrated2$Y, col = rgb(0.5, 0.5, 0.5, 0.4), pch = 19, cex = 0.7,
	xlab = "X1", ylab = "Y", main = "Mechanism 2: Y vs X1", font.lab = 2)
	points(Dat2$data_rct$X1[Dat2$data_rct$A==0], Dat2$data_rct$Y[Dat2$data_rct$A==0], col = "blue", pch = 4, cex = 1.2, lwd=2)
	if(optimal_k_cal2 > 0) points(test_data_cal2$X1[selected_idx_cal2], data_ec_calibrated2$Y[selected_idx_cal2], col = "orange", pch = 19, cex = 0.8)
	legend("topleft", legend = c("Calibrated ECs", "RCT", "Selected ECs"), 
	col = c("gray", "blue", "orange"), pch = c(19, 4, 19), cex = 0.8, bg = "white")
	
	plot(test_data_cal2$X2, data_ec_calibrated2$Y, col = rgb(0.5, 0.5, 0.5, 0.4), pch = 19, cex = 0.7,
	xlab = "X2", ylab = "Y", main = "Mechanism 2: Y vs X2", font.lab = 2)
	points(Dat2$data_rct$X2[Dat2$data_rct$A==0], Dat2$data_rct$Y[Dat2$data_rct$A==0], col = "blue", pch = 4, cex = 1.2, lwd=2)
	if(optimal_k_cal2 > 0) points(test_data_cal2$X2[selected_idx_cal2], data_ec_calibrated2$Y[selected_idx_cal2], col = "orange", pch = 19, cex = 0.8)
	legend("topleft", legend = c("Calibrated ECs", "RCT", "Selected ECs"), 
	col = c("gray", "blue", "orange"), pch = c(19, 4, 19), cex = 0.8, bg = "white")
	
	par(mfrow = c(1, 1))
	dev.off()
\end{lstlisting}

{
Table~\ref{tab:study2} summarizes the performance of different estimators. The na\"ive pooling of all external controls (\texttt{Full}) introduces substantial positive bias ($0.752$). In contrast, the adaptive influence-based estimator ($\hat{\tau}_{\mathcal{S}_{k^*}}$) limits the inclusion of non-comparable samples. It achieves a bias ($0.148$) close to that of the RCT-only AIPW estimator ($\hat{\tau}_{\text{aipw}}$), while substantially reducing the standard deviation from $0.087$ to $0.076$. The calibrated adaptive estimator further improves performance, with bias $0.133$, standard deviation $0.070$, and the smallest MSE among all estimators.
}

\begin{table}[ht]
	\centering
	\caption{Comparison of estimators for Mechanism 2 (True ATE = 3.000)}
	\label{tab:study2}
	\begin{tabular}{lccccc}
		\toprule
		\textbf{Estimator} & \textbf{Estimate} & \textbf{Bias} & \textbf{SD} & \textbf{MSE} & \textbf{$k^*$} \\ 
		\midrule
		Direct ($\hat{\tau}_{\text{direct}}$) & 2.808 & -0.192 & 1.111 & 1.271 & 0 \\ 
		AIPW ($\hat{\tau}_{\text{aipw}}$)     & 3.150 & 0.150 & 0.087 & 0.030 & 0 \\ 
		Full ($\hat{\tau}_{\text{full}}$)     & 3.752 & 0.752 & 0.158 & 0.591 & 400 \\ 
		Selected ($\hat{\tau}_{\mathcal{S}_{k^*}}$) & 3.148 & 0.148 & 0.076 & 0.006 & 110 \\ 
		Calibrated \& Selected & 3.133 & 0.133 & 0.070 & 0.005 & 40 \\ 
		\bottomrule
	\end{tabular}
\end{table}

%

\begin{figure}[htbp]
	\centering
	\includegraphics[width=0.9\textwidth]{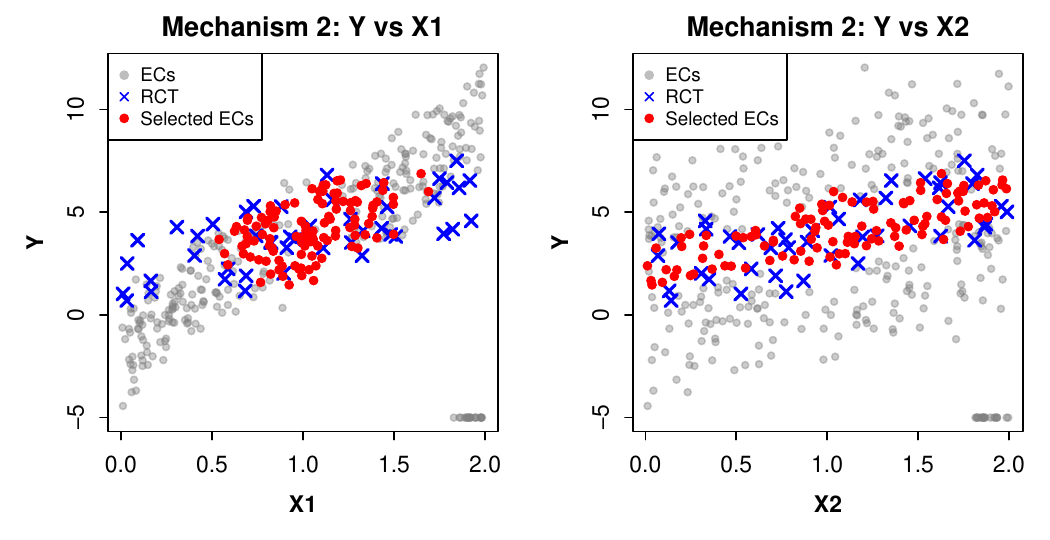}
	\caption{
		The selected external controls (red dots) indicate the subset retained by the adaptive influence-based estimator.}
	\label{fig:mech2_scatter_raw}
\end{figure}

\begin{figure}[htbp]
	\centering
	\includegraphics[width=0.9\textwidth]{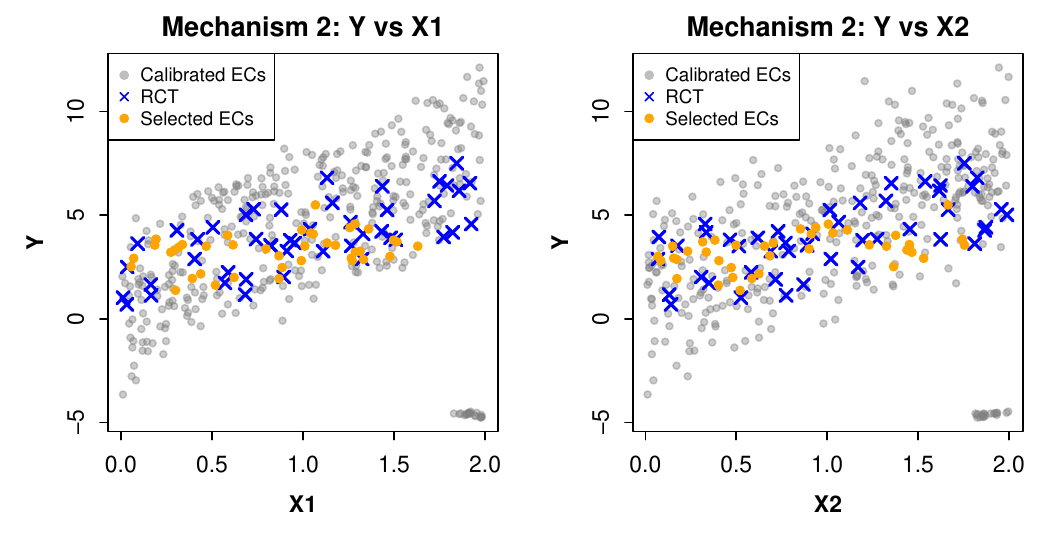}
	\caption{
		The selected external controls (orange dots) indicate the subset retained by the calibrated adaptive influence-based estimator.}
	\label{fig:mech2_scatter_cal}
\end{figure}

{
Figures~\ref{fig:mech2_scatter_raw} and~\ref{fig:mech2_scatter_cal} visualize the external controls and the selected subsets under Mechanism 2 before and after outcome calibration. In the uncalibrated setting, the selected external controls (red dots) avoid regions with strong non-linear distortion and corner outliers. After calibration, the selected subset (orange dots) becomes smaller and more concentrated, reflecting improved compatibility between the external controls and the RCT controls.
}

{
These numerical studies demonstrate that both the adaptive influence-based estimator and its calibrated version provide robust and variance-reduced treatment effect estimates across different structural settings.
}

\subsection{Data Generation Scripts}

{ 
For complete reproducibility of the numerical studies presented above, we provide the R functions used to generate the simulated datasets, along with a helper function for summarizing the estimation results. Specifically, \texttt{gen\_dat\_mech1} and \texttt{gen\_dat\_mech2} specify the covariate distributions, treatment assignment mechanisms, and outcome models (including the systematic bias and deterministic outliers in the external controls), while \texttt{create\_res\_df} is used to organize the reported performance measures for the estimators.
}

\begin{lstlisting}[language=R]
	# =================== Data Generation Functions =================== 
	
	# mechanism 1: 1D covariate, binary logistic outcome
	gen_dat_mech1 <- function(n_rct = 150, n_ec = 600, seed = 123) {
		set.seed(seed)
		n <- n_rct + n_ec
		
		# generate covariates with outliers at the tail
		x <- c(runif(n - 20, 0, 2), runif(20, 1.8, 2.0))
		
		# generate the RCT data
		x_rct <- x[1:n_rct]
		a_rct <- rbinom(n_rct, size = 1, prob = 0.5) 
		y1_rct <- rbinom(n_rct, size = 1, prob = plogis(x_rct + 1))
		y0_rct <- rbinom(n_rct, size = 1, prob = plogis(x_rct - 1))
		y_rct <- (1 - a_rct) * y0_rct + a_rct * y1_rct
		data_rct <- data.frame(X = x_rct, A = a_rct, Y = y_rct)
		
		# generate the EC data (incorporating quadratic bias and outliers)
		x_ec <- x[(n_rct + 1):n]
		a_ec <- rep(0, n_ec)
		prob_y0_ec <- plogis(x_ec - 1 + 2.5 * (x_ec - 1)^2)      
		y_ec <- rbinom(n_ec, size = 1, prob = prob_y0_ec)
		y_ec[(n_ec - 19):n_ec] <- 1 
		data_ec <- data.frame(X = x_ec, A = a_ec, Y = y_ec)
		
		dat_combine <- rbind(data_rct, data_ec)
		dat_combine$R <- c(rep(1, n_rct), rep(0, n_ec)) # R=1 for RCT
		
		return(list(data_rct = data_rct, data_ec = data_ec, dat_combine = dat_combine))
	}
	
	# mechanism 2: 2D covariates, continuous outcome with non-linear EC
	gen_dat_mech2 <- function(n_rct = 150, n_ec = 600, seed = 123) {
		set.seed(seed)
		n <- n_rct + n_ec
		
		# generate 2D covariates with outliers at the corner
		x1 <- c(runif(n - 20, 0, 2), runif(20, 1.8, 2.0))
		x2 <- c(runif(n - 20, 0, 2), runif(20, 1.8, 2.0))
		
		# generate the RCT data (ATE is constantly 3)
		x1_rct <- x1[1:n_rct]
		x2_rct <- x2[1:n_rct]
		a_rct <- rbinom(n_rct, size = 1, prob = 0.5) 
		y0_rct <- 2 * x1_rct + 2 * x2_rct + rnorm(n_rct, sd = 0.5) 
		y1_rct <- y0_rct + 3 
		y_rct <- (1 - a_rct) * y0_rct + a_rct * y1_rct
		data_rct <- data.frame(X1 = x1_rct, X2 = x2_rct, A = a_rct, Y = y_rct)
		
		# generate the EC data (non-linear cubic term and corner outliers)
		x1_ec <- x1[(n_rct + 1):n]
		x2_ec <- x2[(n_rct + 1):n]
		a_ec <- rep(0, n_ec)
		y_ec <- -2 + 4 * x1_ec + 2 * x2_ec + 2 * (x1_ec - 1)^3 + rnorm(n_ec, sd = 0.5)
		y_ec[(n_ec - 19):n_ec] <- -5  
		data_ec <- data.frame(X1 = x1_ec, X2 = x2_ec, A = a_ec, Y = y_ec)
		
		dat_combine <- rbind(data_rct, data_ec)
		dat_combine$R <- c(rep(1, n_rct), rep(0, n_ec))
		
		return(list(data_rct = data_rct, data_ec = data_ec, dat_combine = dat_combine))
	}
	
	# ========= Helper Function for Reporting Table ========
	create_res_df <- function(result_rct, result_full, opt_res_sel, opt_res_cal, true_ate, n_ec) {
		est_dir <- result_rct$estimate["direct"]
		est_aipw <- result_rct$estimate["aipw"]
		est_full <- result_full$estimate
		est_sel <- opt_res_sel$mse_optimal$estimate
		est_cal <- opt_res_cal$mse_optimal$estimate
		
		se_dir <- result_rct$se["direct"]
		se_aipw <- result_rct$se["aipw"]
		se_full <- result_full$se
		se_sel <- sqrt(opt_res_sel$mse_optimal$variance)
		se_cal <- sqrt(opt_res_cal$mse_optimal$variance)
		
		bias_dir <- est_dir - true_ate
		bias_aipw <- est_aipw - true_ate
		bias_full <- est_full - true_ate
		bias_sel <- est_sel - true_ate
		bias_cal <- est_cal - true_ate
		
		mse_dir <- bias_dir^2 + se_dir^2
		mse_aipw <- bias_aipw^2 + se_aipw^2
		mse_full <- bias_full^2 + se_full^2
		mse_sel <- opt_res_sel$mse_optimal$mse
		mse_cal <- opt_res_cal$mse_optimal$mse
		
		res_df <- data.frame(
		Estimate = c(est_dir, est_aipw, est_full, est_sel, est_cal),
		Bias = c(bias_dir, bias_aipw, bias_full, bias_sel, bias_cal),
		SD = c(se_dir, se_aipw, se_full, se_sel, se_cal),
		MSE = c(mse_dir, mse_aipw, mse_full, mse_sel, mse_cal),
		`k*` = c(0, 0, n_ec, opt_res_sel$mse_optimal$top_k, opt_res_cal$mse_optimal$top_k),
		row.names = c('Direct', 'AIPW', 'Full', 'Selected', 'Calibrated & Selected'),
		check.names = FALSE
		)
		return(res_df)
	}
\end{lstlisting}

  \bibliography{bibliography.bib}

\end{document}